\documentclass[useAMS,usenatbib]{mnras}

\pdfoutput=1

\usepackage{mathrsfs}
\usepackage{amsmath,amsfonts}
\usepackage[dvips]{graphicx}
\usepackage{xcolor}
\usepackage{rotating}
\usepackage{txfonts}
\usepackage{dsfont}
\usepackage[T1]{fontenc}
\usepackage{soul}
\usepackage{algorithmic}
\usepackage{algorithm}

\usepackage[multiple]{footmisc}
\usepackage{booktabs}
\usepackage{rotating}
\usepackage{aas_macros}
\newcommand{\mockalph}[1]{}

\def\acena{$\alpha$~Cen~A}
\def\acenb{$\alpha$~Cen~B}
\def\thetav{\boldsymbol{\theta}}
\def\muv{\boldsymbol{\mu}}
\def\epsilonv{{\boldsymbol{\epsilon}}}
\def\fracepscno{\varepsilon_{\mathrm{CNO},c}/\varepsilon_c}

\def\Gammab{{\boldsymbol{\Gamma}}}
\def\Sigmab{{\boldsymbol{\Sigma}}}
\def\Xv{\boldsymbol{X}}

\def\stage{t_{\star}}
\def\alphaov{\alpha_{\mathrm{ov}}}
\def\teff{T_{\mathrm{eff}}}
\def\lzams{L_{\mathrm{ZAMS}}}
\def\lsol{L_{\odot}}

\newcommand{\run}[1]{run~\#{#1}}
\usepackage{color}
\definecolor{colres}{rgb}{0.996,0.945,0.813}

\DeclareMathAlphabet\mathbfcal{OMS}{cmsy}{b}{n}

\def\Exp{\mathrm{E}}

\def\ecno{\varepsilon_{\mathrm{CNO}}}

\def\ec{\varepsilon_{\textnormal{c}}}
\def\epp{\varepsilon_{\mathrm{pp}}}

\def\rcc{r_{\mathrm{cc}}}
\def\reps{r_{\varepsilon}}
\def\alphaov{\alpha_{\mathrm{ov}}}

\title[]{On the uncertain nature of the core of $\boldsymbol{\alpha}$ Cen A}
\author[M.~Bazot]{M.~Bazot$^{1,2}$\thanks{E-mail: mb6215@nyu.edu}, J.~Christensen-Dalsgaard$^{3}$, L.~Gizon$^{1,4,5}$ and O.~Benomar$^{1}$\\
$^{1}$Center for Space Science, NYUAD Institute, New York University Abu Dhabi, PO Box 129188, Abu Dhabi, United Arab Emirates\\
$^{2}$Centro de Astrof\'{\i}sica da Universidade do Porto, Rua das Estrelas, 4150-762, Porto, Portugal\\
$^{3}$Stellar Astrophysics Centre, Department of Physics and Astronomy, Aarhus University, Ny Munkegade 120, DK-8000 Aarhus C, Denmark\\
$^{4}$ Max-Planck-Institut f\"ur Sonnensystemforschung, 37077 G\"ottingen, Germany\\
$^{5}$ Institut f\"ur Astrophysik, Georg-August-Universit\"at G\"ottingen, 37077 G\"ottingen, Germany
}

\begin{document}

\date{Accepted 2011 December 15. Received 2011 December 14; in original form 2011 October 11}

\pagerange{\pageref{firstpage}--\pageref{lastpage}} \pubyear{2002}

\maketitle

\label{firstpage}

\begin{abstract}
High-quality astrometric, spectroscopic, interferometric and, importantly, asteroseismic observations are available for {\acena}, which is the closest binary star system to earth. Taking all these constraints into account, we study the internal structure of the star by means of theoretical modelling. Using the Aarhus STellar Evolution Code (ASTEC) and the tools of Computational Bayesian Statistics, in particular a Markov chain Monte Carlo algorithm, we perform statistical inferences for the physical characteristics of the star. We find that {\acena} has a probability of approximately 40\% of having a convective core. This probability drops to few percents if one considers reduced rates for the $^{14}$N(p,$\gamma$)$^{15}$O reaction. These convective cores have fractional radii less than 8\% when overshoot is neglected. Including overshooting also leads to the possibility of a convective core mostly sustained by the ppII chain energy output. We finally show that roughly 30\% of the stellar models describing {\acena} are in the subgiant regime.

\end{abstract}


\section{Introduction}

{\acena} has long captured the interest of stellar physicists. First, the Centauri system is the closest to our Sun, which means that we have good measurements and a precise picture of the atmospheric parameters of {\acena} (effective temperature, surface chemical abundances, luminosity), but also of its radius, thanks to interferometry. Second, one specific characteristic attracts the attention: it forms a close binary with {\acenb}. This is a critical point, since, over one century, astrometric measurements have given us a good knowledge of the orbit of the system. This in turn provides us with a straightforward way to estimate the mass of {\acena}, which is a critical physical quantity when it comes to modelling a star, maybe the most important one. 

It was not surprising that the first asteroseismic measurements for a solar-like star other than the Sun were obtained for {\acena}. In 2000, two teams observed it almost simultaneously, and obtained the first estimates for {\textquotedblleft}extrasolar{\textquotedblright} stochastic p modes \citep{Bouchy01a,Bedding04}. Since asteroseismology can {\textquotedblleft}drill{\textquotedblright} the depths of stars, these measurements (alongside all the classical ones previously obtained) were used to constrained theoretical models. A particularly interesting problem is that {\acena} has a mass of roughly $\sim$1.1~M$_{\odot}$, the value for which, at solar metallicity, energy starts to be transported by convective motions in the stellar core. A major objective for modelling teams thus became the identification of such convective structures. 

Convective cores are very important features in stellar physics. For intermediate-mass stars (from $\gtrsim 1.7$~M$_{\odot}$ up to $\sim$10-15~M$_{\odot}$), their signatures in isochrones can be a good marker of the presence of overshooting \citep[see e.g.][]{Ribas00}. Other studies have suggested that they could be used as diagnostics for planetary material accretion, and more broadly for planet-formation scenarios \citep{Bazot04,Bazot05}. Some have even speculated on their interactions with dark matter particles \citep{Brandao15}. In the case of {\acena}, given the necessary small size of a potential convective core, the issue is rather to test the limits of astereoseismic data and understand what are the minimal requirements to detect these structures.

An interesting outcome of the first modelling studies on {\acena} was the disagreement that arose between various teams. Some found an optimal model with a convective core \citep{Thevenin02}, some without \citep{Thoul03,Eggenberger04,Miglio05}. They all used the CORALIE seismic data \citep{Bouchy01a}. This discrepancy could be explained by the quality of the data, which would not be constraining enough and allow for multiple local minima in the minimization criteria. It could also stem from the statistical methods used to relate the data and the theoretical models themselves, hence turning the problem into a methodological issue. This has motivated a new observing campaign using the high-precision spectrograph HARPS \citep{Bazot07}. They could identify 34 p modes in good agreement with the previous results. 

These new results were subsequently used by \citet[][hereafter Paper~I]{Bazot12a} in order to estimate the physical characteristics of {\acena}. Their approach differed from those in previous studies in that they applied methodologies of Bayesian numerical statistics to the problem of stellar modelling. The Bayesian approach is focused on the statistical properties of the parameters of the model (considered fixed in the more classical frequentist approach), being given the data (which is the random quantity in the frequentist case). A radical difference with the optimization methods commonly used in stellar modelling is that probabilistic statements about the characteristics of the stars can be made. This brought a new light to the problem of the convective core of {\acena}. In Paper~I, the authors claimed that, being given the data, the odds were roughly $45\%$ that a convective core has developed at the centre of the star.

The study in Paper I was mostly focused on methodological issues, and in particular on comparing stochastic, actually Markov chain Monte Carlo (MCMC), and grid-based sampling strategies for Bayesian statistics in the context of stellar parameter estimation. For that reason, we used a relatively simple version of our stellar evolution code, including for instance the EFF equation of state \citep{EFF}, and limited ourselves to the so-called standard physics (no microscopic diffusion, no extra-mixing besides mixing-length-theory convection, no rotation, no magnetic fields,\dots). In this study we extend the previous analysis to state-of-the-art models. We then we consider the impact of including non-standard physics in our models. In Section~\ref{sect:meth} we present our methodology. After defining the problem, we set our Bayesian statistical model. To that effect we specify the likelihood, the priors on the parameters, but also the stellar model, which relates the physical parameters we are interested in to the statistical objects we manipulate. We describe briefly our sampling strategy. It is an improvement of the simple MCMC approach we used in Paper~I. Our solutions to the estimation problem are presented in Section~\ref{sect:phys}. We interpret them in order to draw a picture of our state-of-knowledge \citep[e.g.][]{Gregory05} on the physical properties of {\acena}, and in particular on the mechanisms for energy production and transport in its core. We also evaluate the impact of changing the nuclear reaction rates, and of including overshoot or microscopic diffusion. Finally, in Section~\ref{sect:blabla}, we discuss briefly two statistical issues related on one hand to the formal description of the parent population of the observed frequencies, and on the other to Bayesian model comparison applied to nuclear reaction rates.

\section{Bayesian estimation}\label{sect:meth}

This paper focuses on the estimation of the physical parameters of {\acena}. This statement contains in itself a major simplification. As already mentioned in the introduction, this star is part of a binary system, a property used to obtain the mass of the star. This also implies several contingent issues. Indeed one should ideally be able to model both stars, {\acena} and B, jointly. In particular they should have similar ages and initial chemical compositions. On one hand this adds to the complexity of the problem since our stellar model has to be generic enough to reproduce observations from objects that may be physically very different. This also reduces the number of parameters to estimate per object. On the other hand, it might help to get more precise estimates for certain parameters, possibly including the age of the star. In the present study, we neglected these issues and focused solely on {\acena}. This was motivated by a technical reason: the algorithm we are using, in its current version, becomes very significantly less efficient when modelling jointly a binary system.

In order to estimate stellar parameters, we chose to use Bayesian Statistics. This was motivated by the fact that their formulation provides a convenient framework for the problem we are facing. Indeed, it is often the case in Natural Sciences that observations differ from measurements obtained from controlled, repeatable experiments. The very nature of the observational apparatuses at play only allow us to acquire data at a given time and position for one or several objects. Such measurements might be subject to changes if taken later, or from elsewhere. Therefore, most of the time, our samples are reduced to one data point (one measurement), that we ought to fit. 

In such cases, one might find difficult to invoke the limit properties of classical statistical tools such as the maximum likelihood estimator. An interesting feature of Bayesian statistics is that it places the parameters at the same conceptual level as the data in classical statistics, i.e. they become random quantities \citep[see e.g.][]{Robert07}. The passage from one description to the other is done thanks to the prior information we have on the parameters (i.e. before getting the data), using Bayes' formula
\begin{equation}\label{eq:bayes}
\displaystyle
  \pi(\thetav|\Xv) = \frac{\pi(\thetav)\pi(\Xv|\thetav)}{\pi(\Xv)}.
\end{equation}
Here $\thetav$ is a vector grouping the parameters of our Bayesian statistical model\footnote{\citet{Robert07} define a Bayesian statistical model as made of a probability density for the data (conditional on the model parameters) and a prior density on the same parameters.} we wish to estimate (see Sect~\ref{sect:ASTEC}) and $\Xv$ a vector containing the data (see Sect~\ref{sect:data}). On the left-hand side, $\pi(\thetav|\Xv)$ is the posterior density function (PDF) of the parameters, the data being given. On the right-hand side $\pi(\thetav)$ is the prior density of the parameters, $\pi(\Xv|\thetav)$ the likelihood (i.e. the probability density of the data, given the parameters) and $\pi(\Xv)$ a normalization constant.

\subsection{Data and likelihood}\label{sect:data}

The first step in order to define our Bayesian statistical model is to define the likelihood function. A very generic approach consists in considering the data as a deterministic average value, $\overline{\Xv}$, plus a stochastic component (i.e. a realization of some randomly distributed noise), $\epsilonv$
\begin{equation}\label{eq:measure_mod}
\Xv = \overline{\Xv} + \epsilonv.
\end{equation}
And we can further write 
\begin{equation}\label{eq:model}
\overline{\Xv} = \mathbfcal{S}(\thetav),
\end{equation}
 with $\mathbfcal{S}$ a mapping from the space of parameters, of elements the $\thetav$, to the space of observables, of elements the $\Xv$. In our case, $\mathbfcal{S}(\thetav)$ symbolizes the stellar evolution codes, which take the components of $\thetav$ as input parameters and produce theoretical, deterministic, estimates of the observations. It can be seen readily that the random variable $\Xv-\overline{\Xv}$ has the same distribution as $\epsilonv$. 

Typical observables in stellar physics are the effective temperature, metallicity-to-hydrogen mass fraction ratio, luminosity, radius, oscillation frequencies. In a more compact form we note them $\Xv = (\teff, Z/X, L, R, \nu_{n,l})$, where the indices $n$ and $l$ run over the observed values for the mode orders and degrees. We will use the same values for the non-seismic variables as in Paper~I; they are reported in Table~\ref{tab:nsobs}. 

In the following, we will make the assumption that the noise is normally distributed for all observables (of course non-identically) and therefore $\pi(\Xv|\thetav)$, $\epsilonv \sim \mathcal{N}(0,\Gammab)$,\footnote{For the sake of compactness, we will sometimes use the symbol {\textquotedblleft}$\sim${\textquotedblright} to indicate that a random variable is distributed according to a specific distribution. By extension we will also use this notation for samples from a random variable to designate the parent distribution.} where $\Gammab$ is the covariance matrix of the data. We make the assumption here that all the components of $\Xv$ are independent and thus $\Gammab = \mathrm{diag}(\sigma_1^2,\dots,\sigma_N^2)$, with $\sigma_i^2$ the variance of the $i$-th observation. 

The likelihood can thus be written
\begin{equation}\label{eq:likeli1}
\displaystyle
\pi(\Xv|\thetav) \propto \exp\left[ -\frac{1}{2} \sum_{i=1}^N \frac{(X_i - \mathcal{S}_i(\thetav))^2}{\sigma_i^2} \right].
\end{equation}
This is a common form for the likelihood often encountered in stellar physics, since it reduces, in an optimization framework, to the least-square minimization method \citep{Eggenberger04,Miglio05,Metcalfe03}. 

Concerning seismic data we focus on the so-called small separations $\delta\nu_{02} = \nu_{0,n} - \nu_{2,n-1}$, which have interesting properties in the framework of stellar modelling \citep{RV03}, rather than the individual frequencies $[\nu]_{n,l}$ themselves \citep{Bazot13}. For the sake of comparison, seismic data will also be taken from \citet{Bazot07}. However, for convenience, and given the new implementation of our MCMC interface, we will use individual small separations, rather than the average one as was done in Paper~I. In practice, this does not seem to change much the final estimation results, the only notable exception being the uncertainties on the age. This might actually make more sense, since it is not always clear how averaging captures the complete behaviour of the small separations\footnote{Even though it was noted by \citet{JCD88} that the knowledge of the average large and small separations is enough to obtain a complete knowledge of the star's evolutionary state, fitting individual separations clearly allows modelling potential higher-order effects of the stellar structure on the frequencies.}.

\begin{table}
\begin{center}
\caption{Classical observational constraints on {\acena}. The last column specifies the reference for the selected quantity.}
\label{tab:nsobs}
\begin{tabular}{@{}lcr@{}}
\toprule
Observable& Value  & Reference\\
\midrule
 $T_{\mathrm{eff}}$& $5810\pm50$~K& \citet{Eggenberger04} \\
 $L/L_{\odot}$& $1.522\pm0.030$& \citet{Eggenberger04}\\
 $R/R_{\odot}$& $1.224\pm0.003$& \citet{Kervella03}\\
 $Z/X$& $0.039\pm0.006$& \citet{Thoul03}\\
 $M/M_{\odot}$& $1.105\pm0.007$& \citet{Pourbaix02}\\
\bottomrule
\end{tabular}
\end{center}
\end{table}

\subsection{Physical model and prior densities}\label{sect:bayesrhs}

\subsubsection{The ASTEC code}\label{sect:ASTEC}

When discussing the physical model we use to account for the observations, $\mathbfcal{S}(\thetav)$, it is necessary to introduce our stellar evolution code. It is indeed a genuine component of the statistical model since, according to Eq.~(\ref{eq:model}), it defines the average value for our measurements in the likelihood. It is through the stellar code that the physical stellar parameters enter the Bayesian statistical model. A brief description is thus in order. 

In the following, we use the Aarhus STellar Evolution Code (ASTEC). It has been described in-depth by \citet{JCD82b,JCD08a}. The oscillation frequencies are computed using the {\tt adipls} module \citep{JCD08b}. There exist various versions of ASTEC and many options. There is a basic setup that is common to all versions we are using in this study and that we refer to as standard physics. It assumes spherical symmetry for the star and neglects the effect of magnetic fields, rotation, wave-induced mixing, mass-loss or radiative levitation. 

Besides those common assumptions, we decided to fix the equation of state and opacities. Regarding the latter, we used the OPAL 95 tables alongside low-temperature values as computed by \citet{Ferguson05}. In addition, we considered the OPAL equation of state in its 2005 version \citep{OPAL02}. This approach uses the so-called \emph{physical picture} to determine the free energy and pressure as cluster expansions \citep[stemming from the application of Feynman diagrams to the many-body problem as suggested by ][]{Bloch58}. A goal of this study is to use the flexibility of ASTEC in order to compare the effect of different input physics on the probability of presence of a convective core, as shall be done in Sect~\ref{sect:phys}. 

There are four main parameters that can be straightforwardly related to physical characteristics of the star. These are the mass of the star, $M$, its age, $\stage$, and those controlling its initial chemical composition, namely the initial heavy-element and hydrogen mass fractions, $Z_0$ and $X_0$.

We then consider two parameters that allow us to model the transport of energy and matter by convection. The first one, rather classical, is the mixing-length parameter, $\alpha$, which characterizes the efficiency  of convection. The mixing-length parameter is implemented as described in the classical parametric convection picture \citep{BV58}. It sets the {\textquotedblleft}mixing-length{\textquotedblright} of the convective flow, that is the characteristic distance over which fluid elements transporting energy travel before dissolving in the medium. Using this simplified picture, it is possible to obtain an analytic expression for the convective energy flux without solving the full set of convection equations. This expression is local and does not take into account possible feedback effects from the bulk of the convection zone. Theory shows that such an assumption is robust when convection is efficient and that the overall stratification is close to adiabatic. 

Such treatment does not however account for potential non-adiabatic effects that may arise, in particular at the limit of the convective zones. The usual Schwarzchild's criterion for dynamical stability merely states that at the neutral point (i.e. where the adiabatic and radiative temperature gradients are equal) the buoyancy forces will not amplify any given departure from the equilibrium radiative stratification. However, it does not tell us that the surrounding medium will brake immediately any fluid element that may reach this point with a given initial velocity gained from convective acceleration \citep[for ampler discussions on the meaning of Schwarzchild's criterion see][and references therein]{Gabriel14}. It is thus natural to expect overshoot phenomena past the boundaries of convective zones, either in the stellar core or envelope. The issue is in fact that there exist no simple theory\footnote{Since, as usual, solving the full set of hydrodynamic equations for convection in the framework of stellar evolution calculations is out of question.} theory to describe such penetration flows. We are thus bound to use \emph{ad hoc} models parametrizing the size of the overshoot layer as
\begin{equation}\label{eq:povs}
\ell_{\mathrm{ov}}  = \alphaov\min(H_p,r_{\mathrm{cc}}),
\end{equation}
 with $\ell_{\mathrm{ov}}$ the size of the overshoot layer, $H_p$ the pressure scale height at the boundary of the convective core and $r_{\mathrm{cc}}$ the radius of the convective core. We have also introduced a parameter, $\alphaov$, which sets the size of the overshoot region in units of the chosen length scale (pressure height or convective core radius). This formulation has been adopted to avoid infinite values of the pressure scale height (as the pressure reaches a maximum)  when the convective core becomes very small. Note that there is also a debate as to whether only chemical elements are mixed in this overshoot layer or if thermalization, leading to adiabatic stratification, also occurs\footnote{Overshooting leading to thermalization is sometimes called penetration, as per the terminology of \citet{Zahn91}.}. Our models use the latter option, even though some evidence \citep{Brummel02,JCD11} seems to favour the former in the case of convective envelopes. 

Finally, we shall mention an important simplification made when dealing with convection in the central regions of stars. We have assumed throughout, including in convective cores, that the nuclear reaction chains ppI, II and III, and the CN cycle are in equilibrium. This may have various implications on the interpretation of our results that will be discussed in Sect.~\ref{sect:neq}.

\subsubsection{Prior densities}\label{sect:priors}

To complete our Bayesian statistical model, we need to specify the prior used in Eq.~(\ref{eq:bayes}) for each stellar parameter. This is a necessary step as otherwise we are only dealing with a likelihood, in which the parameters are not random quantities (and hence cannot be sampled stochastically). Furthermore, the priors depend only on the \emph{a priori} knowledge the practitioner has on the problem at hand (in which range the parameter lies? do we have independent measurements on the parameters?). Therefore, the priors are part of the assumptions made on the Bayesian statistical model, and should be stated as any assumption on any model\footnote{A convenient, albeit simplistic, picture is to consider that priors encode the assumptions we make on our theoretical models (at the level of their free parameters), while the likelihood sums up the assumptions on the data.}.

The first assumption we make is that the priors on the parameters are independent. Effectively this allows us to write
\begin{equation}
\displaystyle
\pi(\theta_1,\dots,\theta_K) = \prod_{k=1}^{K} \pi_k(\theta_k),
\end{equation}
with $\thetav = (\theta_1,\dots,\theta_K)$.

In the following simulations, we will be dealing with five or six free parameters, depending on whether overshooting is considered. The mass has a particular status: since {\acena} belongs to a relatively short-period binary, there exist estimates. We use the value derived by \citet{Pourbaix02}. The assumption made is that their 1$\sigma$-error represent the standard deviation of a normally-distributed random variable. We thus use a (truncated) Gaussian prior on this parameter. It is noteworthy that we chose to use the $1\sigma$ interval given on the mass measurement, instead of the $2\sigma$ or $3\sigma$. This of course may be contended. However, it was already noted in Paper~I that a strong prior on the mass was useful in terms of convergence of the sampling algorithm. Adopting a somewhat optimistic value for the uncertainty on the mass was thus partly guided by numerical considerations.

The case of the age is not so clear. On the one hand, one might argue that there exist priors from gyrochronology. On the other hand, empirical age estimates are usually given on a model-independent scale \citep[see for instance][]{Jeffries14}. This means that a meaningful comparison with the output from ASTEC would require a calibration of that age scale based on our theoretical models. So far, such a work has not been undertaken. We therefore postpone the use of gyrochronology as prior information to future studies and will limit ourselves to uniform priors in this work.

It is also worth mentioning that although we used a uniform prior on the initial hydrogen abundance, it is not a non-informative one\footnote{ Very often uniform priors are used when one does not have very much information on the parameter. It is a way to let it vary freely over a range that is thought to be reasonable. However, one should be careful with such practice. First, the choice of uniform priors as {\textquotedblleft}non-informative prior{\textquotedblleft} can be criticized \citep[see][Sect~3.5]{Robert05}. Second, it appears obvious in many situations that such priors are indeed very informative \citep[see e.g.][and also \citeauthor{Bazot12b} 2012 for a discussion in the context of asteroseismology]{Prosper11}.}. The lower bound has been set after some trial runs. The upper bound, however, was chosen to be consistent with cosmological measurements\footnote{These generally imply the measurement of $^4$He abundances in low-metallicity HII regions and then an extrapolation of the obtained $Z - Y$ relation to the zero-metallicity point (i.e. right after big-bang nucleosynthesis).}. The most recent studies \citep[see e.g.][and references therein]{Olive04,Aver13} have claimed very consistently that the primordial abundance of helium is $Y_p \sim 0.25$ (rounded to the upper digit; good precisions are usually claimed but there exists a certain spread between the published values). Disregarding the abundances of D, $^3$He, and $^7$Li, we consider that the maximum initial hydrogen abundance for {\acena} cannot be higher than $X_0 = 0.75$. The combined priors for $Z_0$ and $X_0$ given in Table~\ref{tab:priors} imply a minimal value for $Y_0$ of 0.246, which is in fair agreement with the estimates of $Y_p$ given in the literature.

The choice of the prior on the initial metallicity and  mixing-length parameter were educated guesses based on commonly-encountered values for these parameters in the literature, atmospheric iron abundances and test runs.

 The choice of the prior on $\alpha_{\mathrm{ov}}$ was more problematic. Most of the previous estimates, usually $\lesssim 0.3$, for $\alphaov$ \citep{Prather74,Maeder81,Ribas00,Cordier03,Demarque04,Dupret04,Zhang12} refer to situations for which $\ell_{\mathrm{ov}} \propto H_p$. This is not guaranteed in our case, since we are looking for a small convective core and it is likely that we will encounter situations for which $\min(H_p,r_{\mathrm{cc}}) = r_{\mathrm{cc}}$. This is coupled to the lack of physical insight we have on the overshoot parameter, the entire model being an \emph{ad hoc} parametrization rather than built from elementary physical principles. We therefore proceeded using a trial-and-error approach in which we explore a vast range of values for $\alphaov$, usually far larger than what conventional wisdom may suggest \citep[although this was already tried for other stars, see e.g.][]{Guenther14}, sampling for models with a central overshoot parameter as large as 2. From these test runs, we noticed that the observations could be reproduced even with extreme values of $\alphaov$ (regardless of the physical meaning of these models). However, we also noticed, while monitoring joint marginal PDFs for $(\alphaov,r_{\mathrm{cc}})$, that $r_{\mathrm{cc}}$ increases linearly with $\alphaov$ up to $\alphaov\sim0.8$. Beyond that, we observed a new regime, with smaller convective cores, not increasing with $\alphaov$. However we also noted that, in this region of the space of parameters, numerical instabilities start to appear in ASTEC. Coupled to the idea, merely guided by intuition, that values of $\alphaov \gtrsim 1$ are unlikely, even when $r_{\mathrm{cc}}$ is the characteristic length scale, we decided to set a prior that would only encompass the first regime. Hence, we decided to use $\pi(\alphaov) = K$, with $K$ a constant when $\alphaov \in [0,0.75]$ and 0 elsewhere. We also tested models without overshoot, in which case $\pi(\alphaov) = \delta(\alphaov)$.

\begin{table}
\begin{center}
\caption{Priors used on the stellar parameters. The second column gives the functional type of the prior, the third the selected parameters of the distribution. For truncated Gaussian, the mean, standard deviation, lower and upper bounds are given. For uniform distributions, the lower and upper bounds given.}
\label{tab:priors}
\begin{tabular}{@{}lcr@{}}
\toprule
Parameter& Prior  & Dist. parameters\\
\midrule
$M$ (M$_{\odot}$)     & Truncated Gaussian         & $[1.105, 7\times10^{-3},1.07, 1.13]$\\
$\stage$ (Gyr)       & Uniform                    & $[1, 8]$\\
$Z_0$                & Uniform                    & $[0.01,  0.04]$\\
$X_0$                & Uniform                    & $[0.6, 0.75]$\\
$\alpha$            & Uniform                     & $[1.2, 2.6]$\\
$\alphaov$          & Discrete/Uniform            & 0/$[0,  0.75]$\\
\bottomrule
\end{tabular}
\end{center}
\end{table}

\subsection{Posterior sampling}\label{sect:MCMC}

Once defined our Bayesian statistical model, we wish to generate $N$ realizations of the stellar parameters $\thetav$, each distributed according to the PDF. Here we denote this sample by $\{\thetav^{(1)},\dots,\thetav^{(N)}\}$, where $\thetav^{(k)}$ is the $k$-th realization. In the case at hand, we write $\{\thetav^{(1)},\dots,\thetav^{(N)}\} \sim \pi(\thetav|\Xv)$ . This can be accomplished using a Markov chain Monte Carlo (MCMC) algorithm. In Paper~I it was argued that these methods should be superior to the classical grid-based approaches often encountered in stellar physics, provided the number of free parameters to estimate is $\gtrsim 5$.\footnote{Note however that, in a grid-based approach, it is not usually necessary to sample time as one can use the (uneven) sampling provided by any code along a stellar evolutionary track. This is not the case for MCMC algorithms, for which a entire sequence of models has to be computed for each required age.} Their basic concept relies on the convergence properties of Markov chains. It is known that, under the assumption of ergodicity, an homogeneous Markov chain will asymptotically produce realizations of random variables all distributed according to a stationary distribution. The idea of MCMC algorithms is thus to produce ergodic Markov chains that have an, a priori unknown, target probability density, in our case $\pi(\thetav|\Xv)$, as their stationary distributions.

Once the sample $\{\thetav^{(1)},\dots,\thetav^{(N)}\} \sim \pi(\thetav|\Xv)$ is generated, it is then possible to use the tools of classical statistics to analyse the distribution of $\thetav$. For instance, we can check for multiple modes, or compute the moments of the distribution in order to provide a useful statistical summary.

In Paper~I the authors used the most generic form of MCMC, that is the Metropolis-Hastings (MH) algorithm, in one of its simplest form \citep{Metropolis53,Hastings70}. The only modification implemented was for the proposal distribution, $q(\thetav^{\boldsymbol{\prime}}|\thetav)$ to be chosen, at each iteration, randomly among three densities (two normal distributions and a uniform one). The proposal distribution controls the exploration of the support of the PDF in the parameter space. The transition kernel, i.e. the probability density for a transition from $\thetav$ to $\thetav^{\boldsymbol{\prime}}$, is $K(\thetav,\thetav^{\boldsymbol{\prime}}) = \rho(\thetav,\thetav^{\boldsymbol{\prime}})q(\thetav^{\boldsymbol{\prime}}|\thetav) + [1-\int\rho(\thetav,\thetav^{\boldsymbol{\prime}})q(\thetav^{\boldsymbol{\prime}}|\thetav)d\thetav^{\boldsymbol{\prime}}]\delta_{\thetav}(\thetav^{\boldsymbol{\prime}})$, with $\delta_{\thetav}(\thetav^{\boldsymbol{\prime}})$ the Dirac mass function and
\begin{equation} 
\displaystyle
\rho(\thetav,\thetav^{\boldsymbol{\prime}}) = \min\left\{ \frac{\pi(\thetav^{\boldsymbol{\prime}}|\Xv)}{\pi(\thetav|\Xv)}\frac{q(\thetav|\thetav^{\boldsymbol{\prime}})}{q(\thetav^{\boldsymbol{\prime}}|\thetav)},1 \right\}.
\end{equation}
The role of $q(\thetav,\thetav^{\boldsymbol{\prime}})$ is thus primordial. In fact, its choice conditions the efficiency of an algorithm; to put it roughly: how fast is reached the asymptotic regime of the Markov chain. This is an issue with very practical and operational consequences. It may indeed be very difficult to set $q(\thetav,\thetav^{\boldsymbol{\prime}})$ so that the MH algorithm is of any use. It was in order to scan a larger range of potential proposal densities that the aforementioned modification was introduced in Paper~I, inspired by a previous work by \citet{Andrieu99}.

However, this approach still proved relatively inefficient, since it required many trials to set properly the two normal distributions in the mixture\footnote{And very often, one of them turned out to be very inefficient.}. Therefore, we improved our strategy by using an Adaptive MCMC algorithm (AM). Let us consider a normal proposal density $q_{(t)}(\thetav,\thetav^{\boldsymbol{\prime}}) = \mathcal{N}(\thetav,\Sigmab^{(t)})$, where $\Sigmab^{(t)}$ is a covariance matrix, that is allowed to change at each iteration. The goal is to adapt the covariance matrix so that an optimal acceptance rate is achieved by the MCMC algorithm. Of course, a Markov chain generated that way will not be homogeneous. Nevertheless, \citet{Haario01} have shown that an algorithm using a normal proposition density with covariance matrix\footnote{It was discussed by \citet{Andrieu08} how AM algorithms could be defined in the broader framework of \emph{vanishing adaption}, which includes the algorithm given by \citet{Haario01}.}
\begin{gather}
\displaystyle
\Sigmab^{(t)} = \frac{1}{t-1} \sum_{i=1}^t (\thetav^{(i)} - \muv_t),\\
\muv_t = \frac{1}{t} \sum_{i=1}^t \thetav^{(i)},\label{eq:movmean}
\end{gather}
still has adequate convergence properties allowing to sample from the target distribution. To ensure the efficiency of the algorithm, the covariance matrix is scaled by a factor $(2.38)^2/d$, with $d$ the dimension of the parameter space \citep[this value was derived using the principles of optimal scaling in MCMC; for an introduction see for instance][]{Brooks12}. Tests with our MCMC/ASTEC interface have shown a significant gain in efficiency, at the level of of both algorithm convergence and supervision, when the AM scheme is implemented.

We shall note here that simple convergence indicators such as the monitoring of (\ref{eq:movmean}) are always satisfactory for the simulations presented in Sect.~\ref{sect:phys}, in the sense that their variation is always well within the credible intervals estimated for the parameters. We do not detect any sign of {\textquotedblleft}non-convergence{\textquotedblright}. However, improvement is still highly desirable. Even though this algorithm performs well in classical tests\footnote{Such as the estimation of the banana-shaped distribution suggested in \citet{Haario01} and other tests \citep[see e.g.][]{Brooks12}.}, it significantly under-performed when applied to stellar models. In particular, we observed acceptance rates approximately four times lower than expected. This of course does not mean that our Markov chains did not converge, but it points out that room still exists to improve the efficiency of the sampling. Finally, it should be mentioned that we run a single Markov chain for each simulation. Beside the impossibility to use some convergence indicators \citep{Gelman92}, this should not differ from running multiple chains. Indeed, the fundamental point is to reach convergence, and one converging chain ought to sample the exact same distribution as multiple converging chains. This was a practical trade-off aiming at gaining some flexibility for the test different physics and/or priors, with regard to the available computing power.

In the following, we will use the posterior densities to provide estimates of physical characteristics of interest. When uncertainties are needed, and due to potential asymmetries in the distributions, we will use the posterior mode, i.e. the value that maximizes the distribution of interest, and associated $100(1-\eta)\%$ credible intervals, which are defined as 
\begin{equation}\label{eq:ci}
C = \{\phi| \pi(\phi|\Xv) \geq q(\eta)\},
\end{equation}
for $\phi$ in the domain of $\pi$, with $0 \leq \eta \leq 1$ and $q(\eta)$ the largest constant such that $P(C|\Xv) \geq 1 - \eta$. In our discussion, we will also often refer to the posterior mean, estimated as the average of our sample from the corresponding posterior density. Considering a random variable $x$, we will denote from now on its posterior mean by $\overline{x}$ and its posterior mode by $\tilde{x}$. We also recall that the probability of a random variable $X$ with density $\pi$ to belong to a set $A$ is
\begin{equation}\label{eq:proba}
\displaystyle
P(X \in A) = \int_{A} \pi(x)dx.
\end{equation}

\section{Physical parameters of {\acena}}\label{sect:phys}

\begin{figure}
\center
\includegraphics[width=\columnwidth]{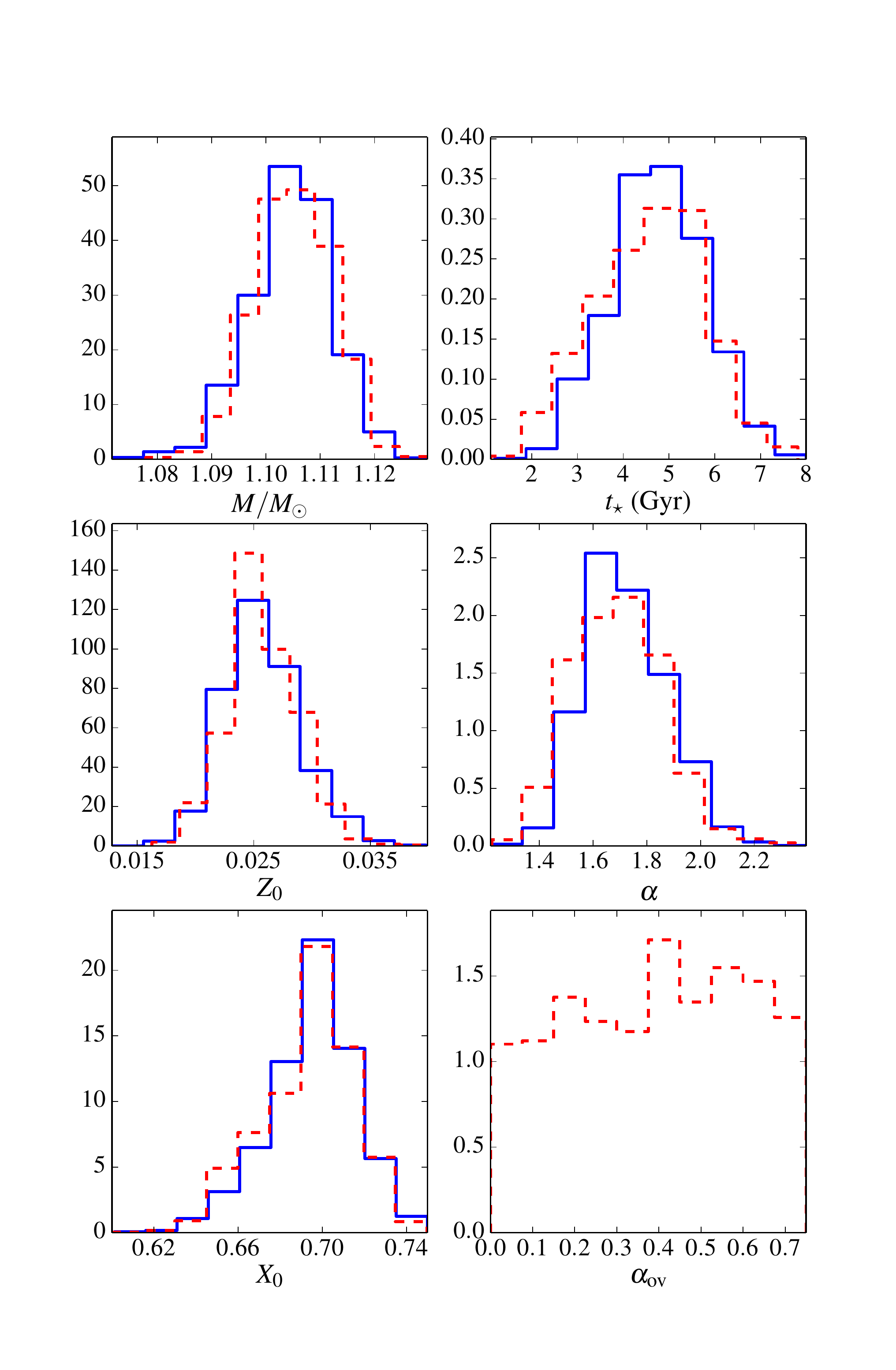}
\caption{Marginal posterior densities for the stellar parameters obtained from runs $\#1$ and $\#2$ in Table~\ref{tab:parameters}.  In each panel are represented the PDFs for the case without (blue, full line) and with overshooting (red, dashed line). In the lower right panel, giving the marginal posterior for the overshoot parameter, only the latter case is represented.}
\label{fig:NACRE_marginals}
\end{figure}

\begin{figure}
\center
\includegraphics[width=\columnwidth]{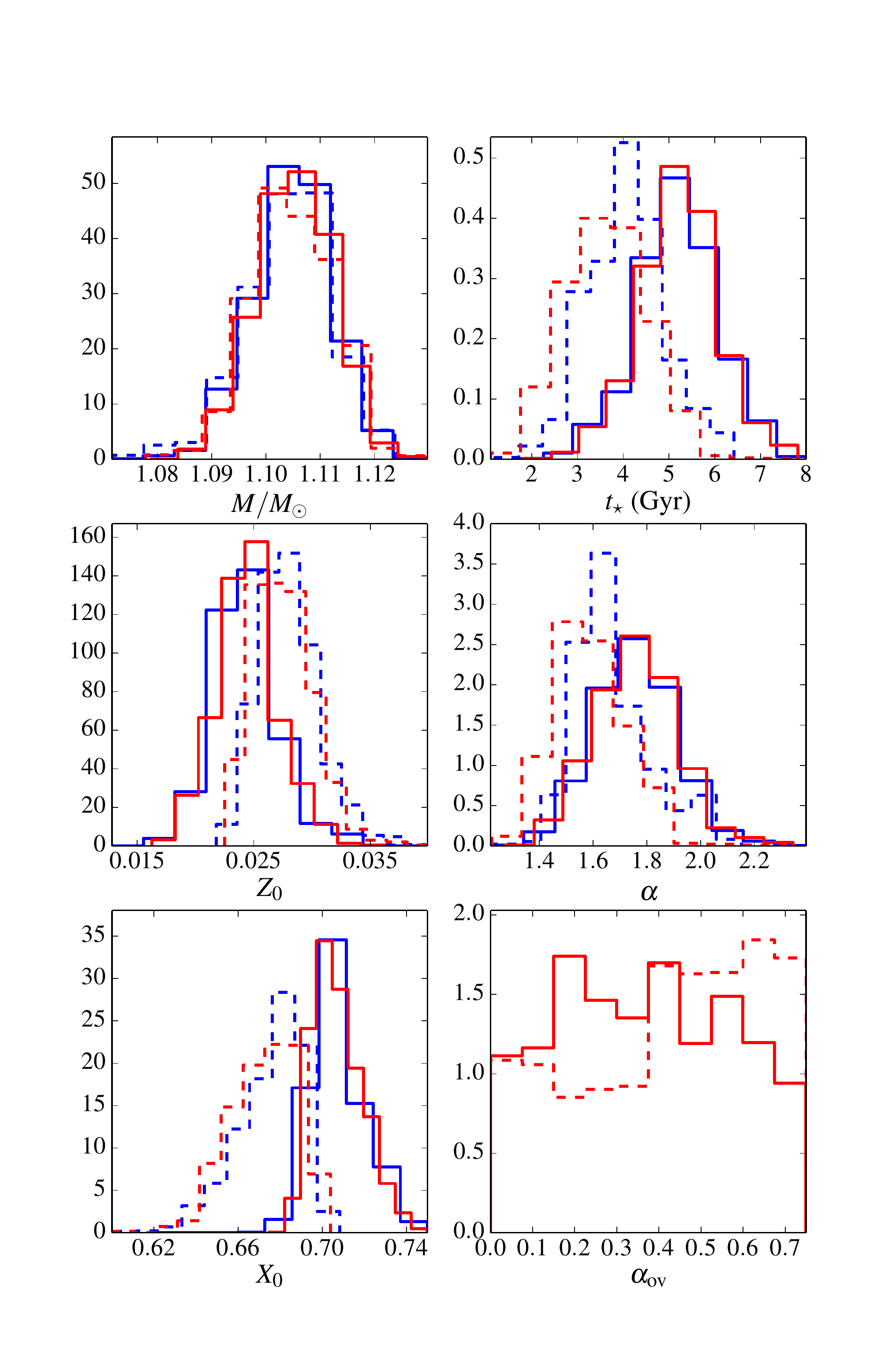}
\caption{Marginal posterior densities for models without (full lines) and with (dashed lines) convective cores. In blue are shown the results from \run{1} and in red from \run{2}.}
\label{fig:NACRE_CC}
\end{figure}

\subsection{Standard physics}\label{sect:base}

For the first cases we studied, we used the NACRE reaction rates \citep{Angulo99}. The prior on $\alphaov$ is $\pi(\alphaov) = \delta(\alphaov)$. It is the setup closest to \run{13} of Paper~I and will serve as a reference model in the following. Table~\ref{tab:parameters} gives the estimates for the stellar parameters and associated credible intervals as defined above. We also show the marginal densities for each stellar parameter in Fig.~\ref{fig:NACRE_marginals}. There are some slight departures from gaussianity in the cases of $\stage$, $\alpha$, $X_0$ and $Z_0$. Our new estimates are very close to those obtained in Paper~I. The only difference of note is the credible interval on the age, which is roughly twice as large here. This is due to the use of the individual small separations rather than the average one as observational constraint. In Paper~I, the uncertainty on the average was estimated after making the assumption that the observed individual small separations were realizations of a single Gaussian random variable $\mathcal{N}(\mu,\sigma^2)$, hence leading to a standard deviation on the mean equal to $\sigma/\sqrt{N}$, with $N$ the number of realizations. Most certainly, this assumption can be questioned, since the small separations are also a function of the frequency, which is not properly taken into account in this case. The newly derived uncertainty on the age might thus be seen as more faithful to our state of knowledge on {\acena}, even though the goal of this paper is not to settle the issue of which seismic indicator is the best, this subject remaining up to debate.

The important question we wish to address in this study is how energy is transported in the central regions of {\acena}. From the MCMC simulations, we can straightforwardly obtain the odds of presence of a convective core. It is simply given by the probability $P(\rcc \neq 0 |\Xv)$, since we can regard the convective core radius as a function of the stellar parameters, $\rcc = \rcc(\thetav)$. Thus obtaining a sample $\{ \thetav^{(1)},\dots,\thetav^{(M)}\}$ also provides us with a sample $\{ \rcc^{(1)},\dots,\rcc^{(M)} \}$. We can also derive, when needs be, the characteristics of the core, its radius and mass. We report the estimated value of $P(\rcc \neq 0 |\Xv)$ and $\rcc$ with its associated credible interval in Table~\ref{tab:parameters}. This physical setup leads to a 37\% chance for {\acena} to have a convective core. The radius of the convective core, in the sense of the posterior mode of the marginal density $\pi(\rcc|\Xv,\rcc \neq 0)$, is $\widetilde{\rcc} = 0.04_{-0.01}^{+0.01}R_{\star}$, with $R_{\star}$ the radius of the star. Again, this is remarkably similar to the estimate from Paper~I, namely $0.04_{-0.08}^{+0.01}R_{\star}$. This leads us to our first conclusion that, in order to model the interior of {\acena}, the details of the formulation of the equation of state or the opacity are not important. For potential effects of these formulations to be observed thanks to seismology, one would require a much higher level of precision. 

It remains to understand the physical conditions under which {\acena} might undergo a convective onset at its centre. In Paper~I, it was argued that $Z_0$ was the critical factor and that over a certain limit, energy generation in the innermost layers would become dominated by the CNO cycle, thus triggering convection. The calculations presented here, which allowed us to study in much greater detail the physical characteristics of the star (in particular due to various improvements in the implementation of the MCMC/ASTEC interface) show that this picture was inaccurate on several accounts, as will be detailed below.

\begin{figure*}
\center
\includegraphics[width=\columnwidth]{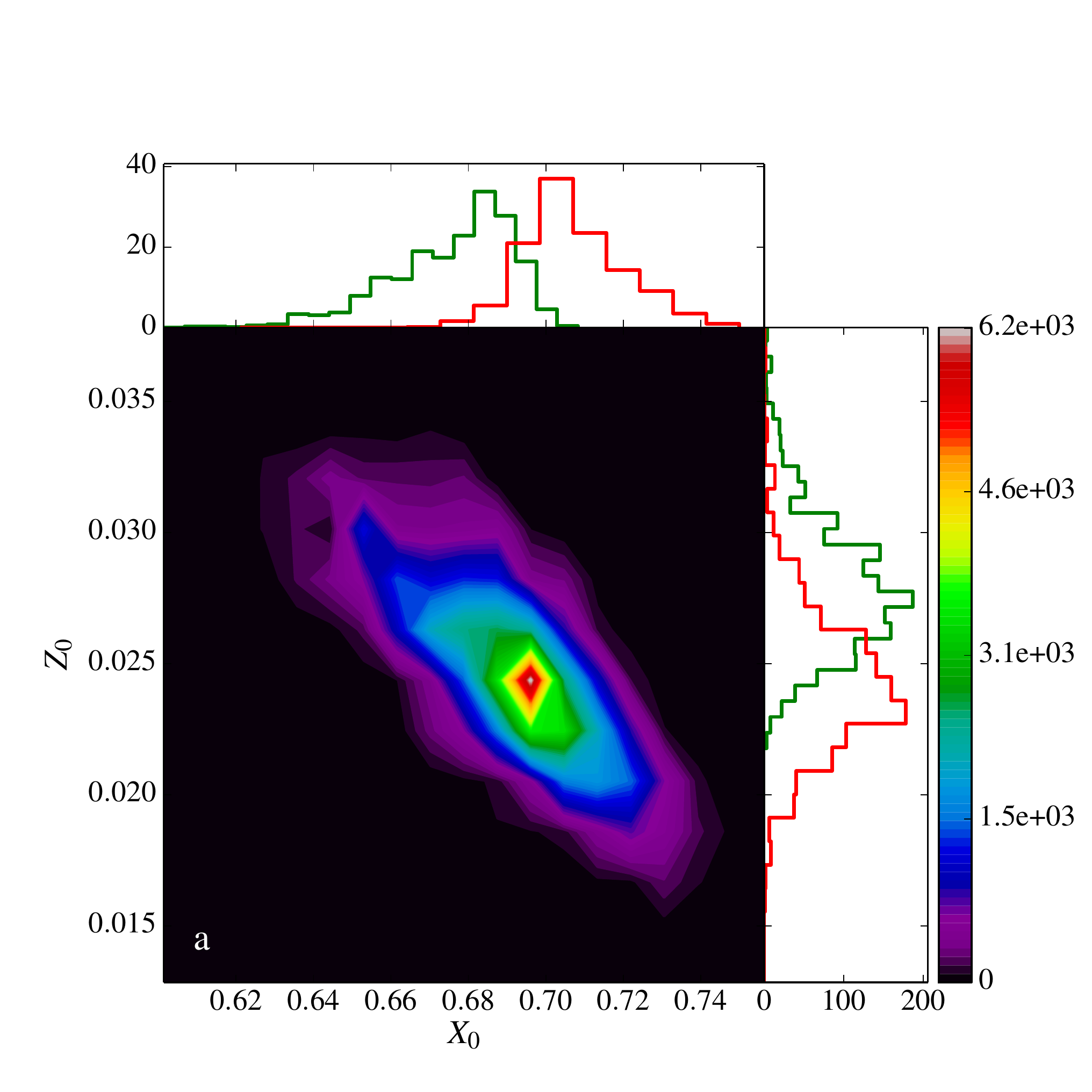}
\includegraphics[width=\columnwidth]{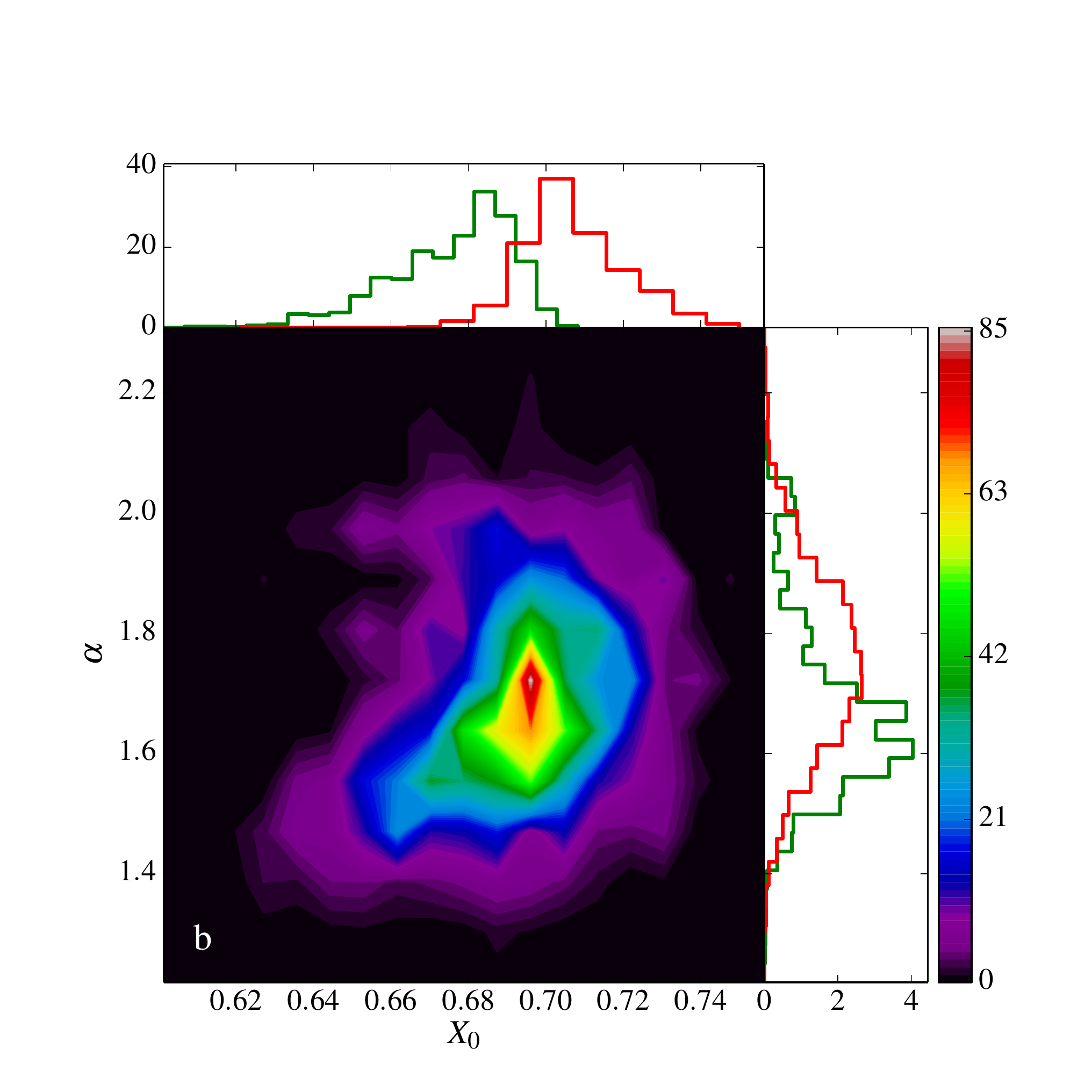}
\includegraphics[width=\columnwidth]{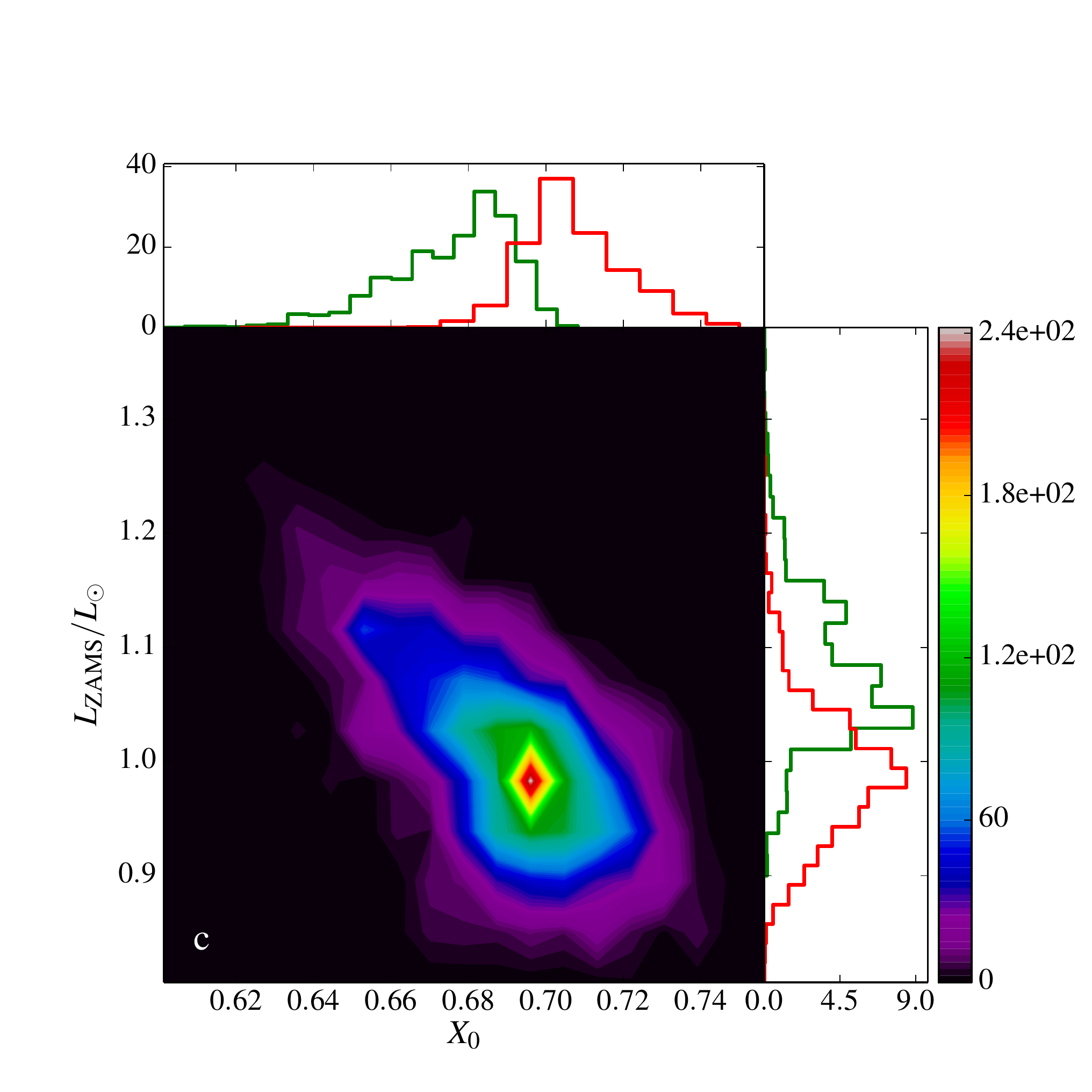}
\includegraphics[width=\columnwidth]{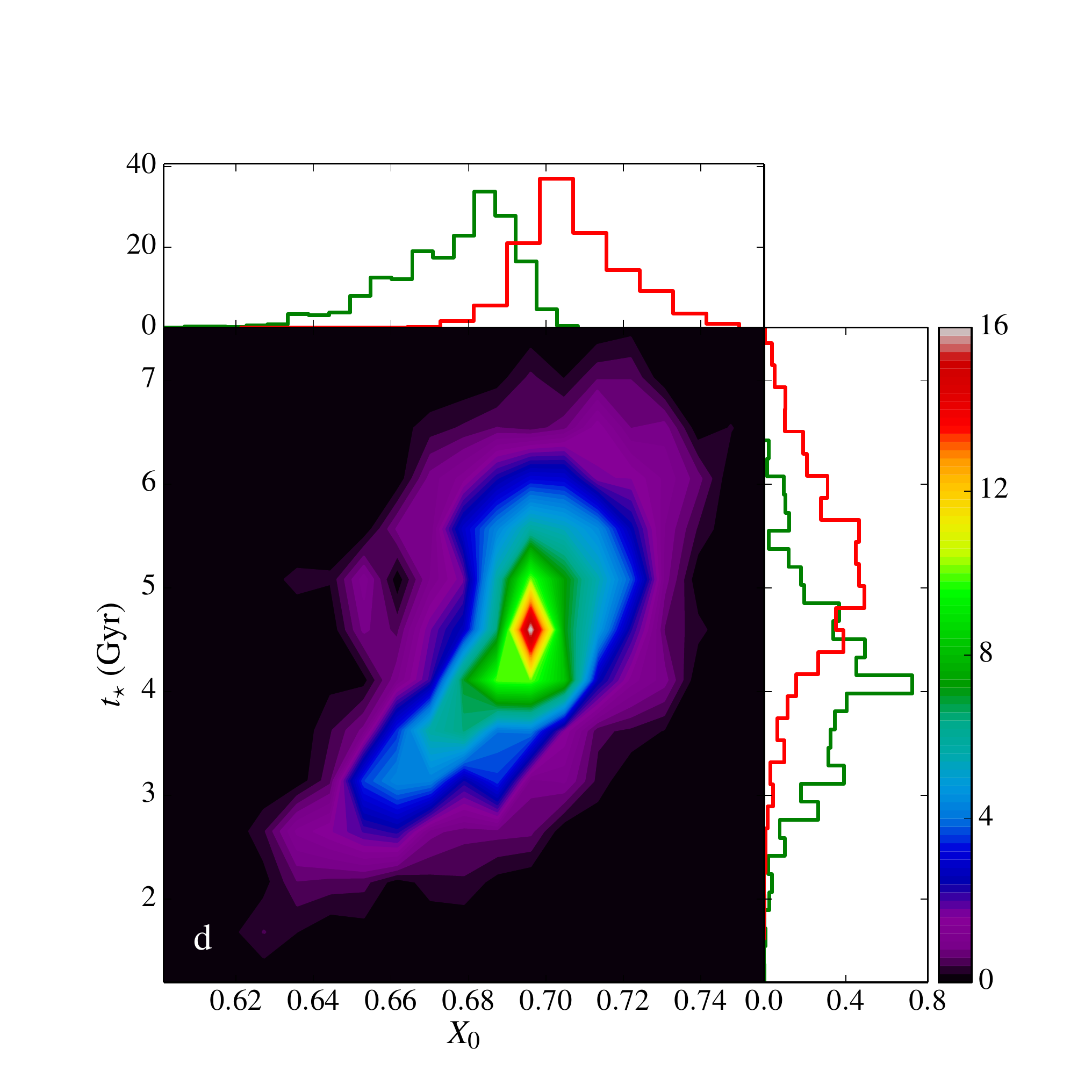}
\caption{Posterior joint PDFs for (a) $X_0$ and $Z_0$, (b) $X_0$ and $\alpha$, (c) $X_0$ and $L_{\mathrm{ZAMS}}/L_{\odot}$ the luminosity of the zero-age main-sequence progenitor of {\acena} in solar luminosity units, (d) $X_0$ and $\stage$. These distributions have been obtained from \run{1}. The smaller side panels display the marginal PDFs for each individual quantity. Two populations are distinguished, in red the radiative-core models and in green the convective-core models.}
\label{fig:jpdfs_NACRE_ovs0_ZAMS}
\end{figure*}

\begin{figure*}
\center
\includegraphics[width=\columnwidth]{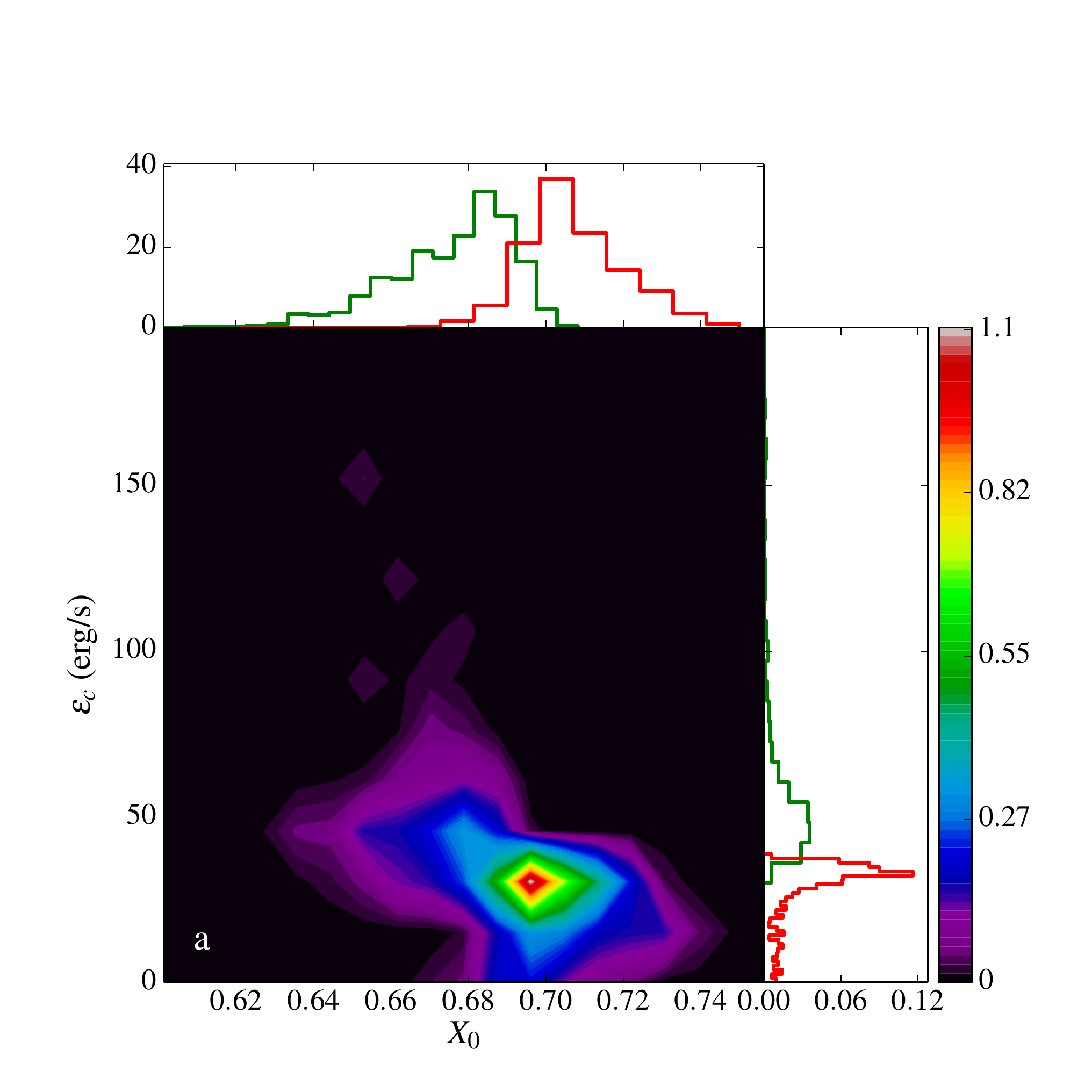}
\includegraphics[width=\columnwidth]{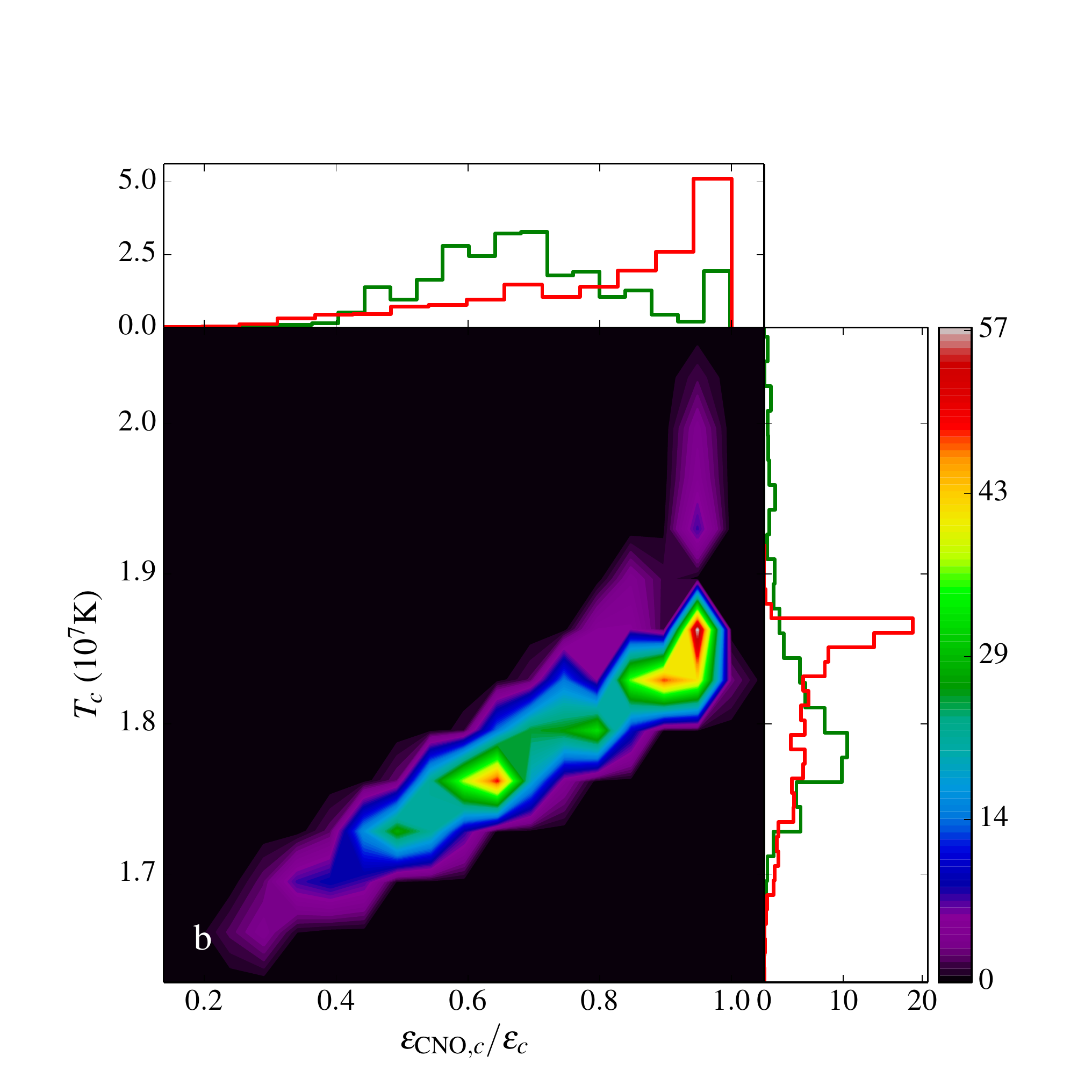}
\includegraphics[width=\columnwidth]{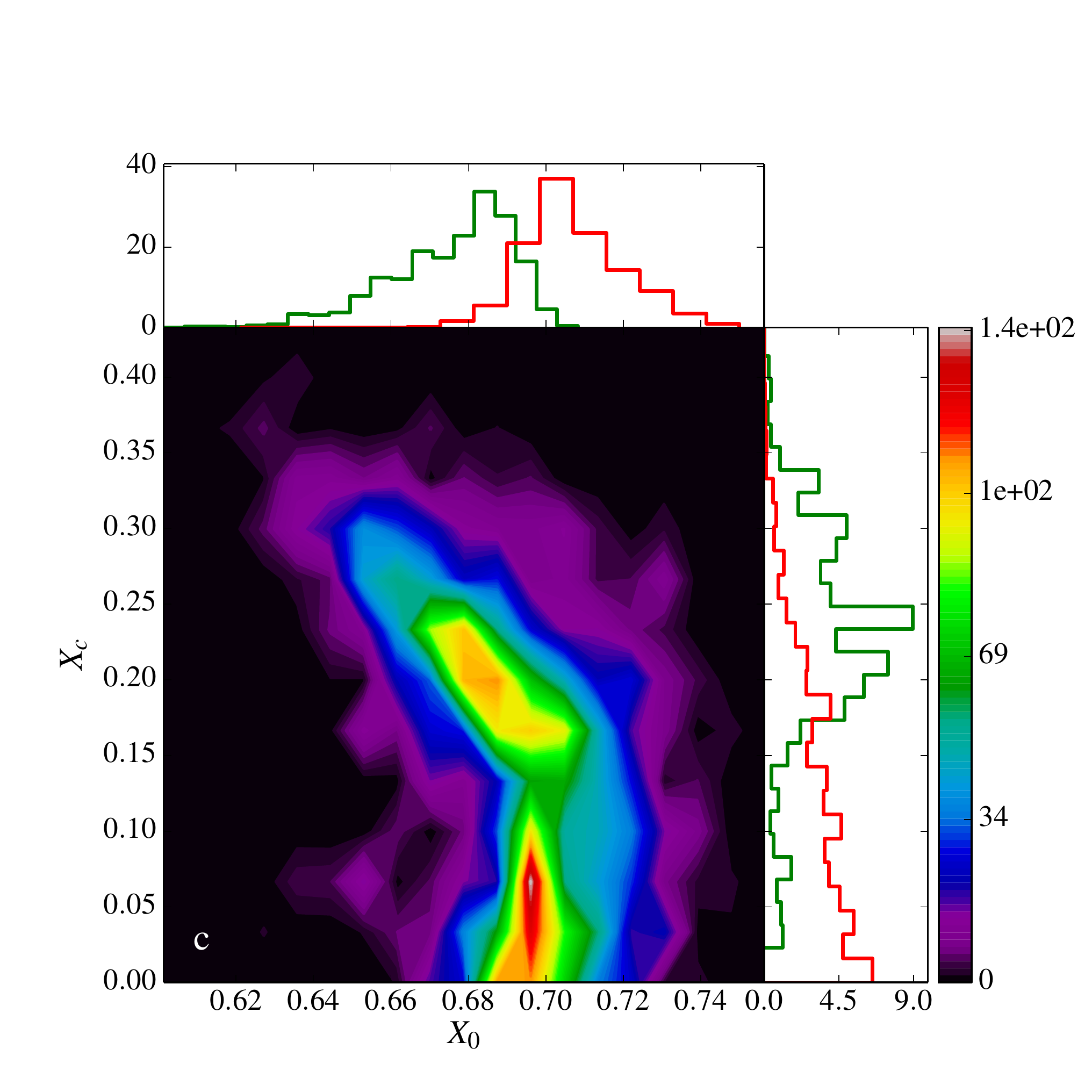}
\includegraphics[width=\columnwidth]{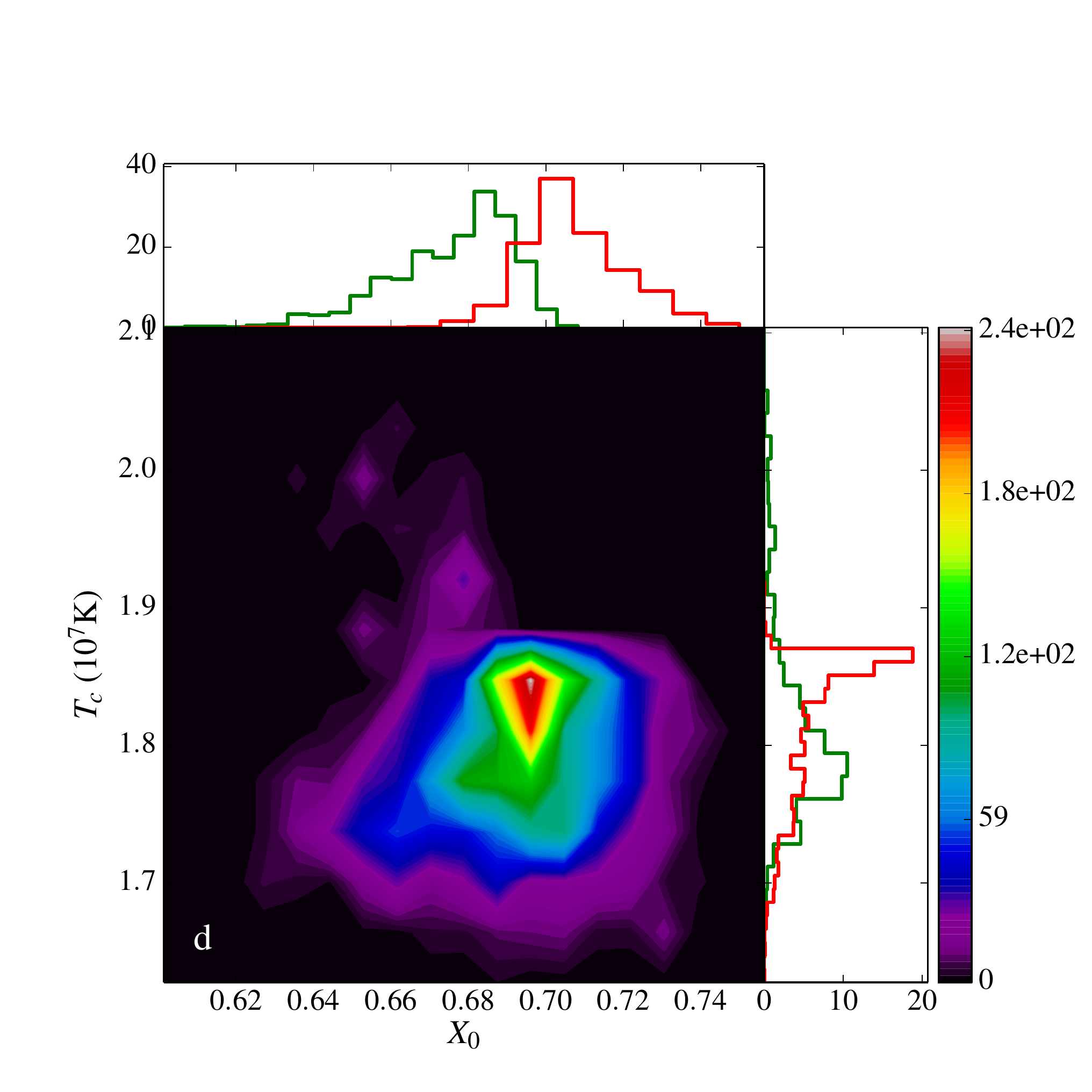}
\caption{Posterior joint PDFs for (a) $X_0$ and $\varepsilon_c$ the central energy generation rate, (b) $\fracepscno$, the relative rate of CNO energy generation rate at the centre, and $T_c$, the central temperature, (c) $X_0$ and $X_c$ the central hydrogen mass fraction, (d) $X_0$ and $T_{c}$. These distributions have been obtained from \run{1}. The smaller panels display the marginal PDFs for each individual quantity. Two populations are distinguished, in red the radiative-core models and in green the convective-core models.}
\label{fig:jpdfs_NACRE_ovs0_centre}
\end{figure*}

A look at Fig.~\ref{fig:NACRE_CC} allows us to distinguish two different families of solutions to the estimation problem, depending on the presence of a convective or radiative core. Due to the strong prior we used on the mass, this parameter has the same behaviour in both regimes. The remaining parameters can be divided the following way: models with convective cores are, on average, more metallic with a lower initial hydrogen mass fraction. They are also younger and their mixing-length parameter is, again on average, lower than for models with radiative cores. One can explain this behaviour by considering two subspaces of the parameter space and their elements, respectively $(X_0,Z_0,\alpha)$ and  $\stage$. The former controls the position on the ZAMS and the value of the latter mostly reflects a necessary adjustment in order to reproduce the physical characteristics of {\acena} (in particular its luminosity), being given these initial conditions. Indeed, increasing $X_0$ roughly amounts to decreasing the luminosity and the effective temperature, and conversely. This could be expected from simple mass-luminosity relations. We can consider homologous stars, as a crude proxy for our stellar models on the ZAMS. In that case, we have\footnote{To obtain this, one can use the reasoning of \citet{Schwarzschild58} and assume an opacity $\kappa \propto Z(1+X)$.} $L \propto \mu^{7.5}(1+X)^{-1}Z^{-1}$. As can be seen in Fig.~\ref{fig:jpdfs_NACRE_ovs0_ZAMS}a, this is compensated, at least partially, by a decrease in $Z_0$, which enhances the luminosity and effective temperature. However, a linear regression with respect to our $\{ (\ln X_0^{(1)},\ln Z_0^{(1)}),\dots, (\ln X_0^{(M)},\ln Z_0^{(M)})\}$ sample shows that a relative increase in $X_0$ corresponds, on average, to a relative decrease in $Z_0$ three times larger in our solution. Intuitively, large enough variations of $X_0$, over the range defined by our prior density, cannot be compensated by a corresponding change in $Z_0$, mostly because this latter is directly constrained by the observations. In order for the ZAMS models to be suitable progenitors for the current epoch {\acena}, the effects of the initial hydrogen mass fraction can be balanced by a simultaneous adjustment of $Z_0$ and $\alpha$. An increase in the latter can increase the effective temperature, everything otherwise equal \citep[see e.g.][]{Clayton84}, while leaving the luminosity relatively unchanged. This can be understood by recalling that the innermost layers of the star, where nuclear energy is produced, are not affected by a change in the convective envelope, which only represents a very small fraction of the overall stellar mass. Therefore, a change in $\alpha$ will not impact the luminosity. However, it will affect the precise location of the radius, in particular because it strongly affects the superadiabatic region near the surface \citep{Gough76,Demarque97}. Based on these simple remarks, we can thus expect that, besides the aforementioned $X_0 - Z_0$ anticorrelation should exist an $X_0 - \alpha$ correlation. This seems indeed to be the case, as seen in Fig.~\ref{fig:jpdfs_NACRE_ovs0_ZAMS}b. We can also see in Fig.~\ref{fig:jpdfs_NACRE_ovs0_ZAMS}c that $X_0$ and $\lzams/\lsol$ anticorrelate, with $\lzams$ the luminosity on the ZAMS. This can be explained straightforwardly by considering that the cumulative effects of $Z_0$ and $\alpha$ balance more fully the impact of $X_0$ on the effective temperature than on the luminosity. From there on, the age difference between radiative and convective-core models is relatively simple to explain. The low-$X_0$ models have lower $\lzams$ and therefore need more time to reproduce both $L$ and the seismic constraints. It is thus a simple evolutionary effect (it is indeed observed, although not plotted here, that $\stage$ anticorrelates with $\lzams/\lsol$). Moreover, the mass of {\acena} being decorrelated from all the other stellar parameters, we do not expect it to impact the characteristic evolutionary timescale.

We now need to relate the stellar parameters $\thetav$ to the central conditions of {\acena} in order to understand the driving mechanisms behind potential convective onsets. In Fig.~\ref{fig:jpdfs_NACRE_ovs0_centre}a, we show the joint posterior density for $(X_0,\varepsilon_c)$, with $\varepsilon_c$ the central energy generation rate. We first note that there is a relatively clear-cut threshold at $\epsilon \sim 40$~erg/s over which the star has a convective core. We do not observe, either for radiative or convective-core models, any clear correlation between $X_0$ and $\varepsilon_c$. This contrasts with the results of Paper~I and may be due to sampling issues in the previous study. We consider the present results to be more reliable.

\begin{figure}
\center
\includegraphics[width=\columnwidth]{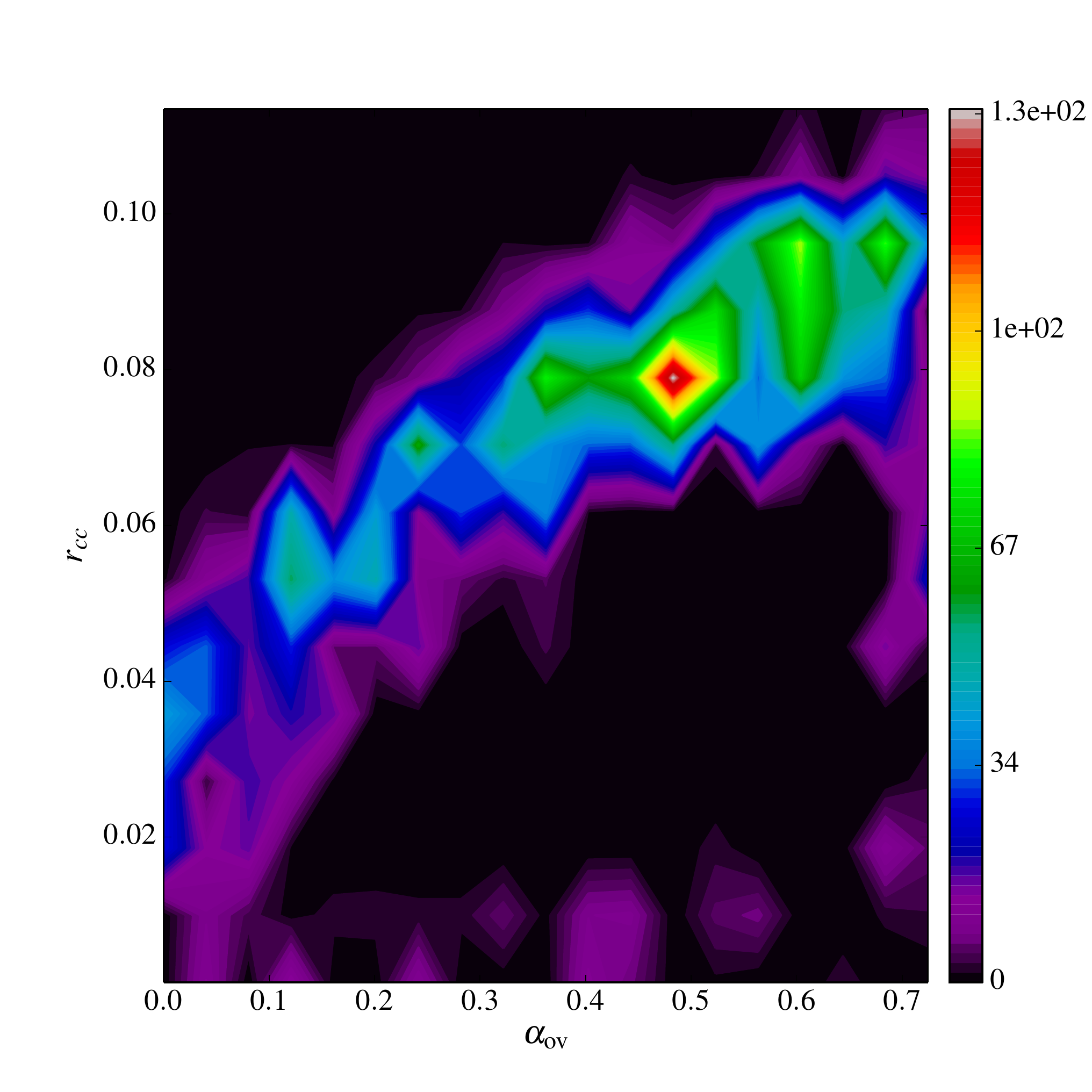}
\caption{Joint PDF for $\alphaov$ and the radius, $\rcc$, of the convective core in stellar radius units.}
\label{fig:alov_rcc_NACRE}
\end{figure}

Figures \ref{fig:jpdfs_NACRE_ovs0_centre}b, c and d give keys to understanding the core of {\acena}. There we show the posterior joint densities for, respectively, $(T_c,\fracepscno)$, $(X_0,X_c)$ and $(X_0,T_{c})$, where $T_c$ is the central temperature of {\acena}, $\varepsilon_{\mathrm{CNO},c}$ and $X_c$ are the central energy generation rate due to the CNO cycle and hydrogen mass fraction, respectively. The first striking observation comes from Fig.~\ref{fig:jpdfs_NACRE_ovs0_centre}b, which shows the weight of the CNO cycle in the overall energy generation process at the centre of the star. Even though the posterior means of $\fracepscno$ for radiative and convective-core models, respectively 0.80 and 0.68, are relatively close, the shapes of the marginal densities $\pi(\fracepscno|\Xv,\rcc=0)$ and $\pi(\fracepscno|\Xv,\rcc\neq0)$ tell an interesting story. In fact the former one increases steadily from $\sim$0.3 up to 1, where it reaches its maximum. It indicates that the there is no one-to-one correspondence between central convective onset and the CNO cycle taking over the pp chain. To set ideas straight, two-third of the radiative-core models have $\fracepscno > 0.75$ and a quarter of them $\fracepscno > 0.95$. In contrast, $\pi(\fracepscno|\Xv,\rcc\neq0)$ peaks close to its mean value and much fewer models involve very high proportions of energy generation by the CNO cycle (less than 10\% convective-core models have $\fracepscno > 0.95$). 

To understand this, one has to disentangle once more the effects of the initial chemical composition and evolution. During the main sequence, in this mass range, there is a transition in the energy generation process. On the ZAMS, the pp chain is dominant. It progressively becomes less and less important as the central temperature rises due to an increase of the mean molecular weight. This is valid for models with both radiative and convective cores. In order for the latter to develop, the stars shall have a central temperature that is high enough when the central hydrogen mass fraction $X_c$ is still significant. The second part of this requirement is related to the dependence of the pp and CNO cycles on $X$, with $\epp \propto \rho X^2$, with $\epp$ the energy generation rate from the pp chain, and $\ecno \propto \rho XX_{\mathrm{CNO}}$.\footnote{$X_{\mathrm{CNO}}$ is taken here, according to a popular approximation, as $\propto Z$. In our case the proportionality coefficient is 0.249} It is because this condition is not fulfilled that radiative-core models may have higher central temperatures and $\fracepscno$, as seen in Fig.~\ref{fig:jpdfs_NACRE_ovs0_centre}b, than convective-core ones. 

Finally, Fig.~\ref{fig:jpdfs_NACRE_ovs0_centre}d gives an interesting picture of the effect of evolution on the central temperature. On the ZAMS, the correlation between $T_c$ and $X_0$ (not shown here) reflects closely the behaviour of $\lzams$, seen in Fig.~\ref{fig:jpdfs_NACRE_ovs0_ZAMS}c. However, for the current-epoch {\acena} evolution has blurred the distinction between radiative and convective-core models. We can nevertheless see in the left panel of Fig.~\ref{fig:jpdfs_NACRE_ovs0_centre}d that $\pi(T_c|\Xv,\rcc=0)$ reaches a maximum at $\sim$1.9$\times10^7$~K and drops suddenly above this value. The absolute value of this peak's location is determined by the remaining central hydrogen mass fraction. If this latter is low and the CNO cycle is dominant in the central regions, then the temperature sensitivity of this nuclear burning process does not allow for models too hot to be selected (in order to reproduce the luminosity of {\acena}). This is the thermostating effect already discussed in Paper~I.

\begin{figure*}
\center
\includegraphics[width=\columnwidth]{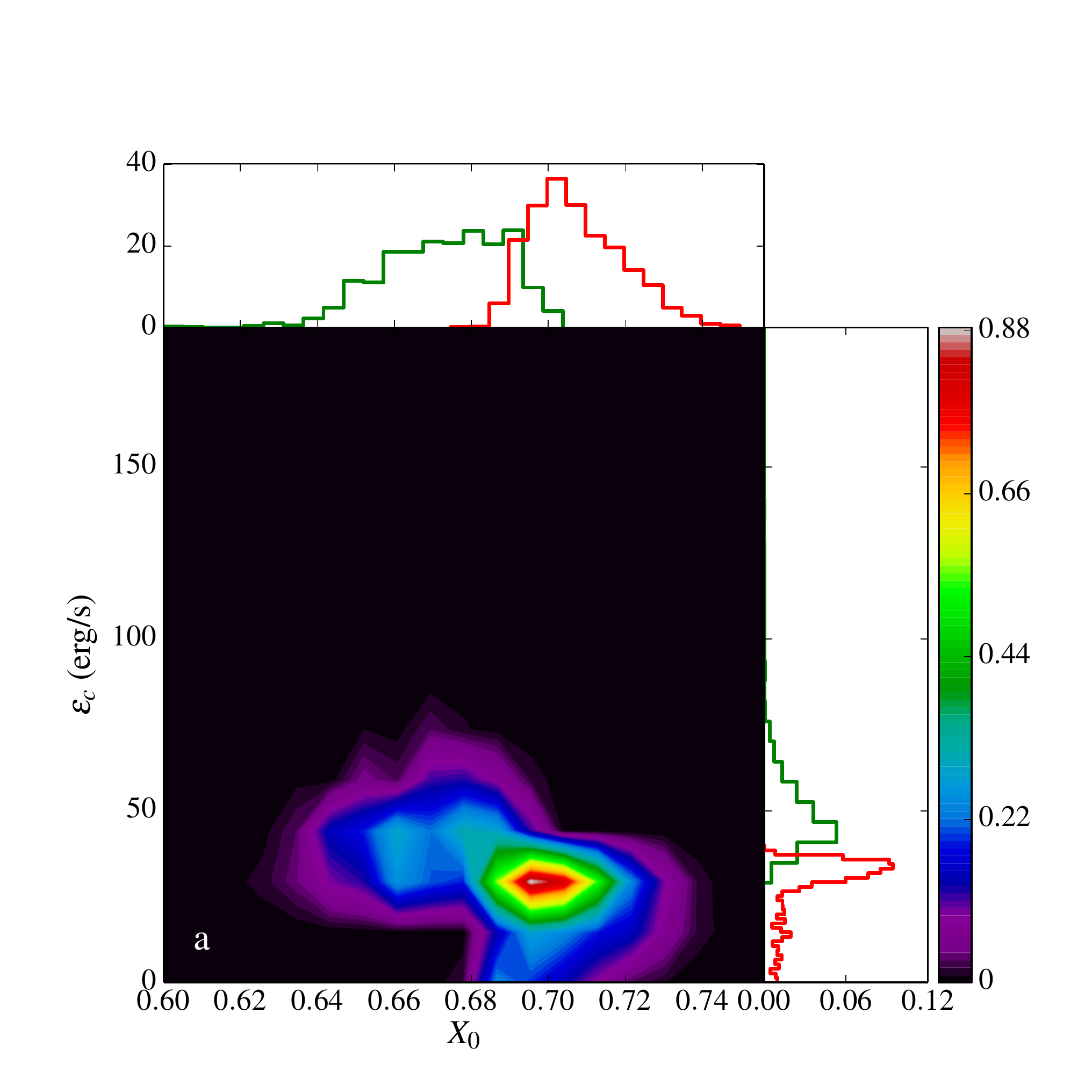}
\includegraphics[width=\columnwidth]{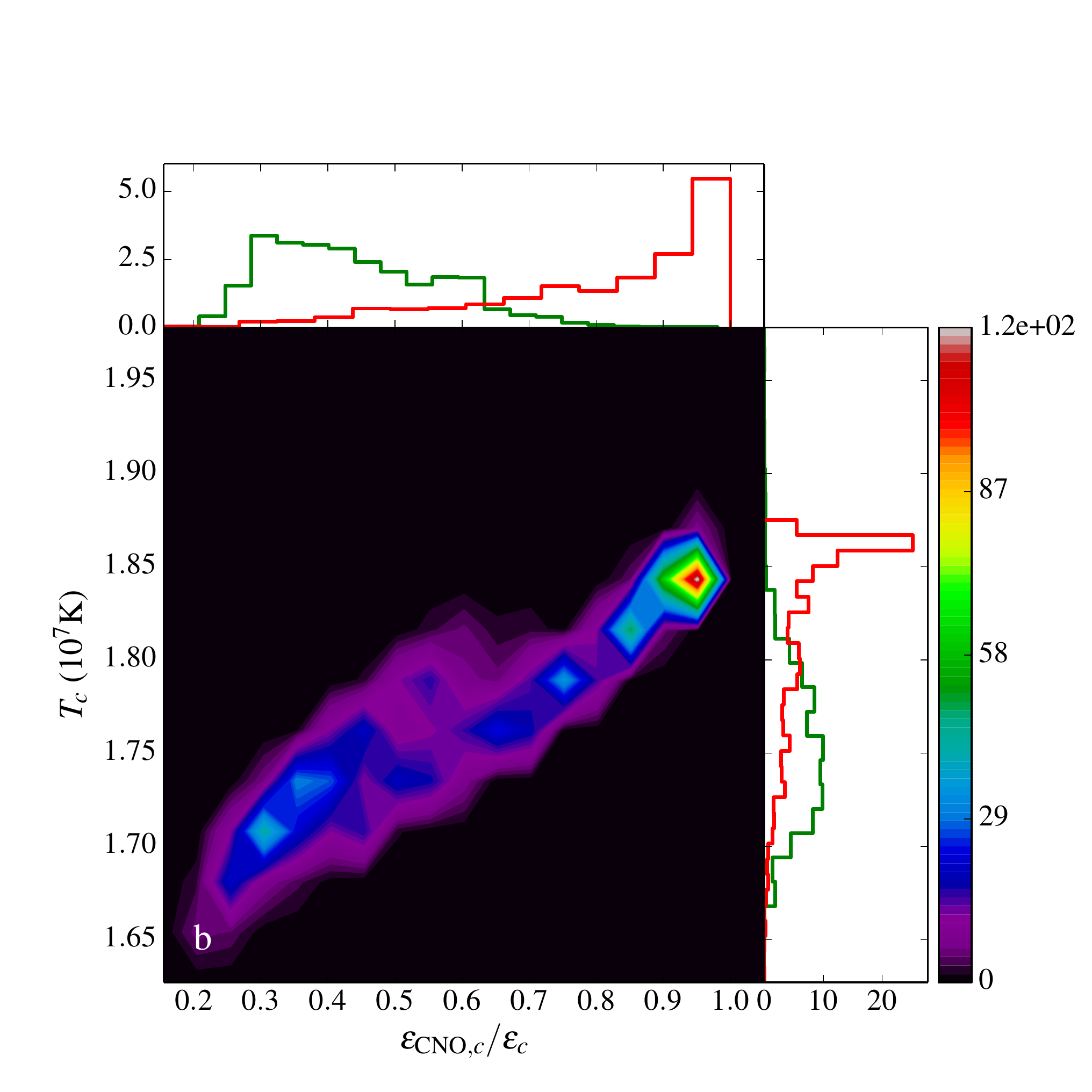}
\includegraphics[width=\columnwidth]{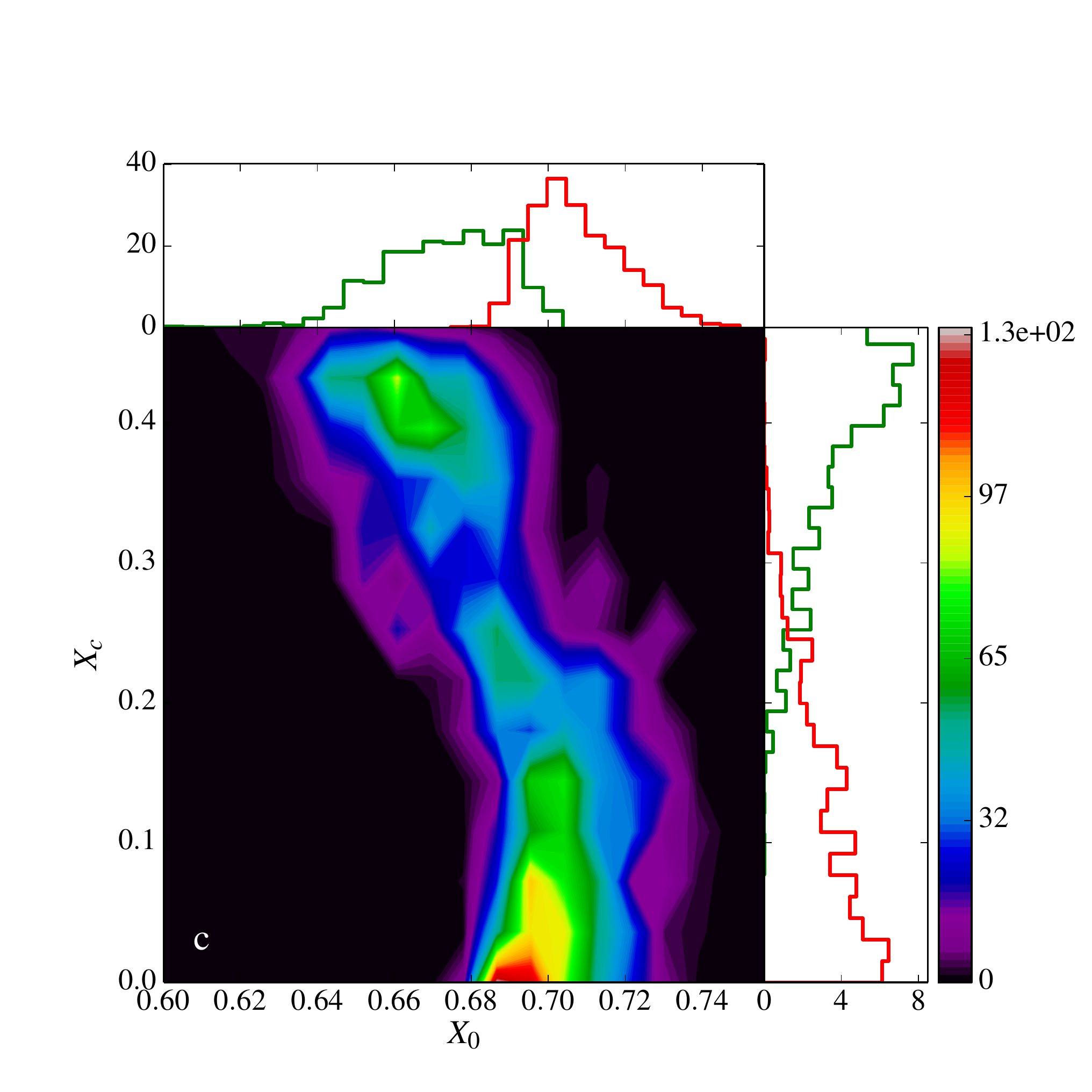}
\includegraphics[width=\columnwidth]{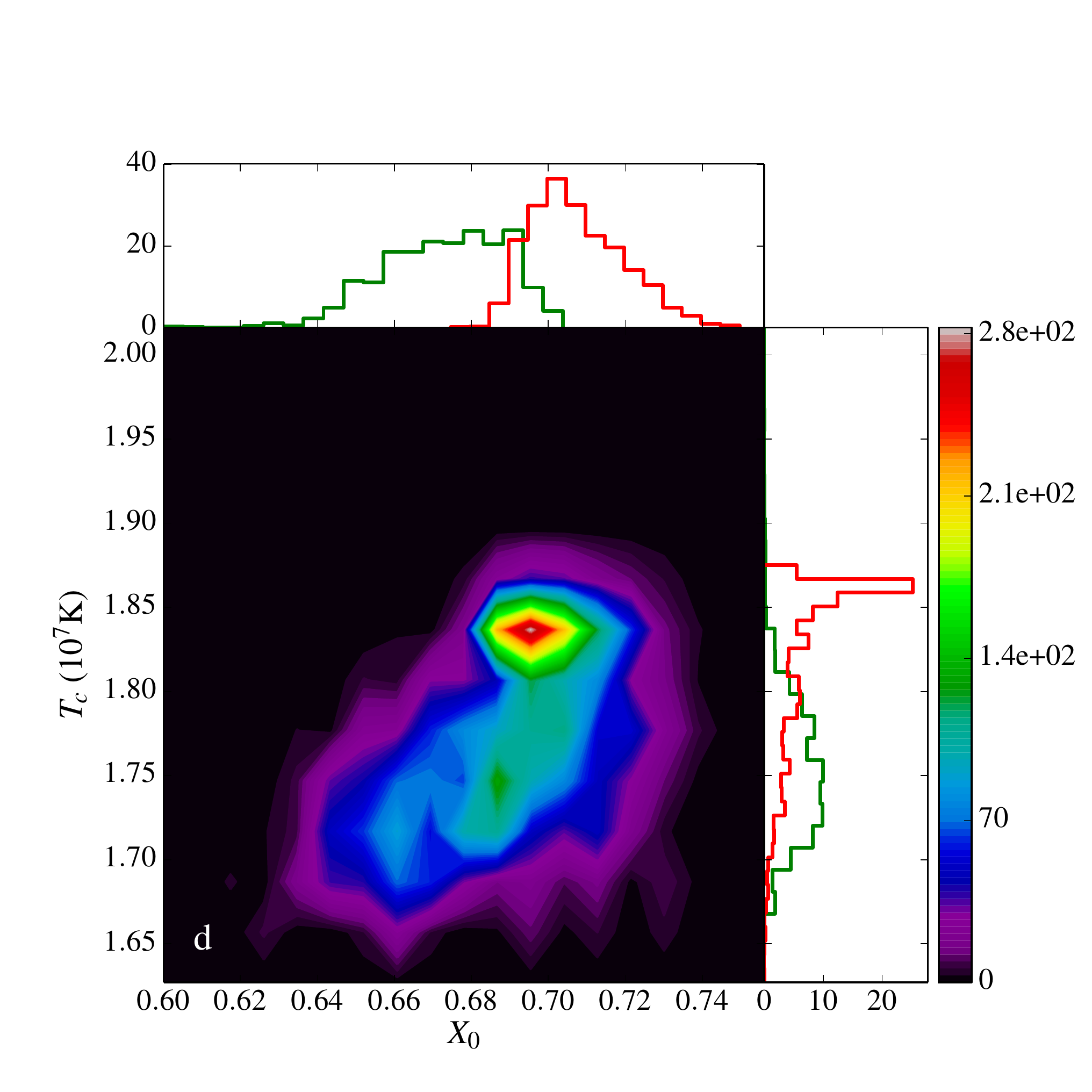}
\caption{Same as Fig.~\ref{fig:jpdfs_NACRE_ovs0_centre} but for the results of \run{2}.}
\label{fig:jpdfs_NACRE_ovs075_centre}
\end{figure*}

 This aspect should also be linked to the tail of $\pi(\varepsilon_c|\Xv,\rcc=0)$ that we observe in Fig.~\ref{fig:jpdfs_NACRE_ovs0_centre}a. It indicates that for a non-negligible proportion of radiative-core models, the energy production at the centre becomes extremely low. Further inspection shows that some radiative-core models with low central hydrogen mass fractions (to fix ideas, we obtained $P(X_c < 0.1|\Xv, \rcc=0) = 0.33$ and $P(X_c < 0.01|\Xv, \rcc=0) = 0.04$) also have very low luminosities over a small extent beyond $r=0$, thus implying a nearly isothermal core since $l\sim dT/dr$. This is perhaps best seen by looking at the location of the peak of energy production in the stellar interior $\reps = \mathrm{argmax}[\epsilon(r)]$. It turns out that $P(\reps\neq0|\Xv,\rcc=0) = 0.52$. Roughly half the radiative-core models for {\acena} are in the very early stage of the shell-burning stage, i.e. already on the subgiant branch. The total probability for {\acena} being a (very young) subgiant can be obtained using the product rule $P(\rcc = 0, \reps\neq0|\Xv) = P(\reps\neq0|\Xv,\rcc=0)P(\rcc=0|\Xv) = 0.32$. We should stress that potential isothermal cores are in any case at a very {\textquotedblleft}embryonic{\textquotedblright} stage, since an estimate of $\reps$ in the sense of the posterior mean gives $\overline{\reps} = 0.02R_{\star}$.

\subsection{Effect of overshoot}\label{sect:ovs}

The introduction of overshooting adds some complexity to this picture of {\acena}. The overshoot parameter is now free to vary between 0 and 0.75. From Fig~\ref{fig:NACRE_marginals}, the first noteworthy conclusion is that our results favour a star with overshoot. The estimate given in Table~\ref{tab:parameters} for \run{2} is, as expected, larger than values usually encountered in the literature, which we attribute to the fact that the characteristic length scale of the mixed region, set by Eq.~(\ref{eq:povs}) is $\rcc$, the radius of the convectively unstable region, and not $H_p$. Furthermore, we see in Fig.~\ref{fig:NACRE_CC} (lower-right panel) that the distribution of $\alphaov$ is relatively flat for radiative-core models. This means that there is no privileged value that helps to reproduce the current epoch {\acena}. Since the properties of the convective cores are affected by the amount overshoot introduced, this means that, before reaching this point, either the star never had a convective core, or if it had, it was not big enough to leave an observable signature in the final distribution for $\alphaov$. Hence including overshoot does not change the results for radiative-core models and in particular the proportion of them having already entered the shell-burning phase, for these we obtained $P(\reps\neq0|\Xv,\rcc=0) = 0.51$.

The distribution $\pi(\alphaov|\Xv,\rcc\neq0)$ for convective-core models increases slightly with $\alphaov$. This reinforces the conclusion that some mixing process is necessary to better reproduce the data. Nevertheless, it turns out that the only changes induced by this new physical process concern solely the deepest layers of {\acena}, and hence seem to be mostly demanded by the need to reproduce the seismic data. It indeed appears from Figs.~\ref{fig:NACRE_marginals} and \ref{fig:NACRE_CC}, and Table~\ref{tab:parameters}, that the inclusion of overshoot-driven mixing does not modify the results of the estimation problem for the other parameters. In particular, our previous conclusions on the dependence of the solution on $X_0$, $Z_0$ and $\alpha$ remain unchanged. This means that our conclusions on the behaviour of the ZAMS models, and the subsequent evolutionary effects controlling the age of the star, are the same as in the case whithout overshoot.

The main impact of this extra mixing is on the innermost structure on the star, which therefore appears to be poorly coupled to the atmospheric conditions, despite the seismic data. As seen in Fig.~\ref{fig:alov_rcc_NACRE} there is some latitude left to obtain larger convective cores in {\acena}. It is largely confirmed by the fact that including overshooting does not affect much the proportion of convective-core models (see Table~\ref{tab:parameters} ; incidentally this also means that the overall proportion of radiative-core shell-burning models remains unchanged with the inclusion of overshoot). The increase of $\rcc$ is achieved thanks to the overshoot parameter and there is a clear correlation between the two. This geometric effect in turn brings in more hydrogen to the central regions. Although the shape of $\pi(X_0,\varepsilon_c|\Xv)$, seen in Fig.~\ref{fig:jpdfs_NACRE_ovs075_centre}a, does not vary much relative to Fig.~\ref{fig:jpdfs_NACRE_ovs0_centre}a, the distribution for $X_c$ is very different for convective-core models (Fig.~\ref{fig:jpdfs_NACRE_ovs075_centre}c), the central hydrogen mass fraction being around $\sim$0.4 for relatively similar ages. This leads to a drastic change in terms of the energy-generation mechanism. We see from Fig.~\ref{fig:jpdfs_NACRE_ovs075_centre}b that the distribution of $\fracepscno$ now peaks for much lower value than in the overshoot-free case. The posterior mean estimate gives 0.44. This implies that for a significant number of models, the dominant source of energy responsible for convective onset is not the CNO cycle but the pp chain. In fact, $\sim$40\% of convective-core models have $\fracepscno < 0.4$ and only $\sim$25\% have $\fracepscno > 0.5$. More precisely, it is the ppII chain that is the main contributor (unsurprisingly since it is the most energetic branch of the pp process). It is always responsible for more than 80\% of the pp energy generation rate, while the ppI branch contributes less than 20\%. The ppIII branch amounts, on average, to less than 1\% of the total energy generation rate.

The major factor influencing convective-core models when overshoot is present is the in-flow of hydrogen at the centre. As seen in Fig.~\ref{fig:jpdfs_NACRE_ovs075_centre}b the central temperature is in the range $T_7 \sim 1.65 - 1.85$, with $T_7 = T/1\times10^{7}\mathrm{K}$ which means that it lies, in the $(T,\varepsilon)$ plane, close to the transition between pp and CNO-dominated energy production rates. An immediate consequence of the enhanced hydrogen abundance is that the $\epp$ is increased more than $\ecno$ due to their dependence in $X$. Since the two rates become roughly equal in this temperature range, then an increase of $X$ might lead to a dominance of the pp over the CNO cycle. Following \citet{Parker64} we recall that $\epsilon_{\rm pp} \propto \psi \rho X_{\rm H}^2$, where $\psi$ is a parameter reflecting the dominating branch of the pp chains and approaching 2 when the ppII branch dominates. Under the conditions relevant to the present study $\psi$, as a function of temperature, is close to its maximum, with $\psi \sim 1.6 - 1.8$. Therefore, overshoot allows for enough hydrogen and helium to be present in the core of {\acena} for the ppII branch to be close to its maximum efficiency, which explains the particular configuration we observe here. 

We can cross-check the validity of this interpretation by looking at the temperature sensitivity of the energy generation rates for convective and radiative-core models, which can be expressed as $\partial\log\varepsilon/\partial\log T$. The former group of models has an average value, taken at maximum energy generation rate, $\overline{\partial\log\varepsilon/\partial\log T(\reps)} \sim 10.\pmb{9}$ and the latter $\overline{\partial\log\varepsilon/\partial\log T(\reps)} \sim 14.\pmb{6}$. These values are significantly higher than those of a pure pp chain in equilibrium, which is usually in the range $\approx 3.5 - 6$. This indicates that the temperature dependence is already dominated by the CNO cycle and explains the thermostating effect seen in Fig.~\ref{fig:jpdfs_NACRE_ovs075_centre}b (right side-panel). Indeed, we see that the distribution peaks at $T_c \sim 1.85\times10^7$~K and then is abruptly cut. The high value of the sensitivity ensures that the temperature cannot become too high in order to respect the luminosity constraint. To complete the description, we note that this thermostating effect accordingly corresponds to the appearance of isothermal cores in centrally radiative models.

The posterior mean of $\overline{\partial\log\varepsilon/\partial\log X}$ is $\sim$1.2 for radiative-core and $\sim$1.4 for convective-core models, as would be expected when the pp chain dominates ($\partial\log\varepsilon/\partial\log X = 2$) the CNO cycle ($\partial\log\varepsilon/\partial\log X = 1$). The latter value is also an indication of the sensitivity of the energy generation rate to the in-flow of hydrogen induced by overshooting and how the ppII chain can be the main driver of a convective core.
 
It is important to note here that this analysis does not mean that the pp chain can sustain alone a convective core. It rather means that given the proper chemical composition, it is possible for the ppII branch to be the main energy provider of the H-burning process while reaching the minimum rate necessary for central convective onset, which we estimate at roughly $\sim$40~ergs/s. This is a remarkable configuration that departs from the traditional picture in which the appearance of a central convective core is usually linked to a dominant CNO cycle. The latter is still very important as will be discussed in the following section. This result shows nevertheless that the continuous transition between the two branches, pp and CNO, allow for a certain variety of configurations. 

\subsection{LUNA reaction rates}\label{sect:LUNA}

We also wanted to assess the influence of the $^{14}$N(p,$\gamma$)$^{15}$O reaction rate on the convective core. This reaction is the bottleneck of the CNO cycle and is involved in its two main branches. It thus controls the evolutionary timescale of CNO-dominated stars. It also is a relatively energetic reaction  ($Q \sim 7.3$~MeV) and a decrease of the associated reaction rate shall lead to a decrease of the overall energy generation rate. 

 Retaining only the first terms in the expansion of the average reaction rate, we recall that the NACRE collaboration found \citep{Angulo99}

\begin{multline}\label{eq:rrt_NACRE}
N_A \langle \sigma v \rangle_{gs} \approx 4.83\times10^7 T_9^{-2/3}(1- 2T_9 + 3.41T_9^2 - 2.43 T_9^3)\\
\times\exp\left[{-15.231T_9^{-1/3} - \left( \frac{T_9}{0.8} \right)^2}\right],
\end{multline}
with $\langle \sigma v \rangle_{gs}$ the average reaction rate at ground-state and $N_A$ the Avogadro number.
 
\begin{table*}
\caption{Estimated parameters of {\acena} from different MCMC simulations. Column~1 gives the index of the model, column~2 the nuclear reaction rates used in ASTEC, column 3-8 give the estimates as the posterior modes of the PDFs for the mass, age, initial metallicity and hydrogen mass-fraction, mixing-length and, if included, overshoot parameter. The upper and lower bounds of associated credible intervals estimated using (\ref{eq:ci}) are also given. Column 9 shows the odds for central convective onset. Column 10 gives the estimated fractional radius of the convective core of {\acena}, if present.}
\label{tab:parameters}
\begin{tabular}{lccccccccc}
\toprule
Run \# &  Nuclear reaction & $M$ (M$_{\odot}$)& $\stage$ (Gyr)& $Z_0$ & $X_0$ & $\alpha$ & $\alphaov$ & \% Convective cores & $r_{\mathrm{cc}}/R_{\star}$\\
\midrule
1 & NACRE & $1.106_{-0.008}^{+0.008}$ & $4.7_{-1.0}^{+1.2}$ & $0.025_{-0.003}^{+0.004}$ & $0.70_{-0.02}^{+0.02}$ &$1.68_{-0.13}^{+0.19}$ &  - & 37 & $0.038_{-0.013}^{+0.011}$\\[.15cm]
2 & NACRE & $1.105_{-0.007}^{+0.009}$ & $4.9_{-1.5}^{+1.0}$ & $0.025_{-0.003}^{+0.004}$ & $0.70_{-0.02}^{+0.02}$ & $1.70_{-0.20}^{+0.16}$ & $0.51_{-0.35}^{+0.15}$ & 41 & $0.084_{-0.025}^{+0.020}$\\[.15cm]
3 & LUNA & $1.106_{-0.008}^{+0.007}$ & $4.7_{-0.8}^{+0.9}$ & $0.027_{-0.004}^{+0.003}$  & $0.69_{-0.02}^{+0.02}$ &$1.71_{-0.13}^{+0.13}$ &   - & 3 & $0.026_{-0.012}^{-0.012}$ \\[.15cm]
4 & LUNA & $1.105_{-0.007}^{+0.007}$ & $4.7_{-0.9}^{+1.0}$ & $0.027_{-0.004}^{+0.004}$ & $0.69_{-0.02}^{+0.02}$ & $1.70_{-0.13}^{+0.14}$ & $0.30_{-0.16}^{+0.30}$ &  2 & $0.079_{-0.037}^{+0.027}$ \\[.15cm]
5 & NACRE & $1.105_{-0.007}^{+0.008}$ & $4.3_{-0.8}^{+0.8}$ & $0.028_{-0.004}^{+0.004}$ & $0.68_{-0.02}^{+0.02}$ & $1.86_{-0.15}^{+0.13}$  &  - & 89 & $0.052_{-0.011}^{+0.008}$ \\[.15cm]
6 & NACRE & $1.105_{-0.008}^{+0.007}$ & $5.4_{-1.2}^{+0.2}$ & $0.025_{-0.002}^{+0.003}$ & $0.70_{-0.02}^{+0.02}$ & $1.77_{-0.17}^{+0.05}$  & - & 37 & $0.038_{-0.011}^{+0.008}$  \\[.1cm]

\bottomrule
\end{tabular}
\end{table*}

The results from the LUNA experiment led to a lower astrophysical factor $S$ \citep{Formicola04}, which in turn implied a revised rate for $^{14}$N(p,$\gamma$)$^{15}$O \citep{Imbriani05}

\begin{multline}\label{eq:rrt_LUNA}
N_A \langle \sigma v \rangle_{gs} \approx 3.12\times10^7 T_9^{-2/3}(0.782- 1.5T_9 + 17.97T_9^2 - 3.32 T_9^3)\\
\times\exp\left[{-15.193T_9^{-1/3} - \left( \frac{T_9}{0.486} \right)^2}\right].
\end{multline}

It was noted earlier by \citet{Magic10} that, \emph{caeteris paribus}, a lower average collision rate for this reaction would reduce significantly the chance of convective energy transport in the innermost stellar layers. They applied these new results to the cluster M67, showing that given two models consistent with the observations, the one with LUNA reaction rates will not have a convective core. We want to test how such a revision impacts our previous estimates. We should also note that there exists another revision of the NACRE reaction rate for $^{14}$N(p,$\gamma$)$^{15}$O \citep{Adelberger11}. We chose not to use it here, even though this will be considered in future studies. They also give a lower reaction rate compared with the results of \citet{Angulo99}.

We ran similar simulations as those described in Sects.~\ref{sect:base} and \ref{sect:ovs} simply switching the reaction rate from (\ref{eq:rrt_NACRE}) to (\ref{eq:rrt_LUNA}). In Table~\ref{tab:parameters}, we give the results for \run{3} and $\#4$, which are the counterparts of, respectively, \run{1} and $\#2$, using LUNA reaction rates. 

First we note that including the LUNA reaction rates indeed change the age of the radiative core models, which are those dominated by the CNO cycle. The posterior mean  estimate of the age for these models is roughly 15\% lower with LUNA. However, this remains well whithin the associated credible intervals. 

Most importantly we see that, as expected, the use of LUNA reaction rates brings down the odds of a convective core developing in {\acena}. The drop is quite spectacular since only 3\% to 4\% of the models have now undergone central convective onset. This also illustrates our statement from Sect.~\ref{sect:ovs} that the CNO cycle, even when not the dominant process plays an important role in energy production. From the output of the MCMC simulations, we can indeed evaluate that the pp chain in runs \#3 and \#4 is responsible for, on average, 9.7~erg/s and 13.8~erg/s, whereas it accounts for 9.9~erg/s and 10.5~erg/s in runs \#1 and \#2. It is thus the difference in $\ecno$ that causes the almost complete disappearance of convective-core models from our solutions. On average, the LUNA reaction rates cause a decrease of $\sim$9~erg/s of $\overline{\varepsilon}$ relative to the NACRE values, the bulk of which is related to the lower CNO energy output. We also checked that the posterior means for the central temperature are the same in runs \#3 and \#4 and runs \#1 and \#2, it is indeed always close to $1.8\times10^7$~K. From these considerations, we shall conclude that, regardless of it being the main energy production process in {\acena}, the CNO cycle is certainly one of the major source of uncertainty, through the nuclear reaction rates, when one wishes to model this star.

One should also note that the similar odds for convective core presence in \run{3} and \#4 confirms our conclusion limiting the role of overshooting to a geometrical one. It simply favours the growth of a convective core but has no bearing on its onset.

Finally, we see that the inclusion of LUNA reaction rates impacts also slightly the picture we had of the radiative-core models. Their initial metallicity is now higher and their initial hydrogen mass fraction lower. On average they are younger than when NACRE reaction rates are used. This reflects an extension of the {\textquotedblleft}permissible{\textquotedblright} region in the parameter space for radiative-core models. Otherwise, the relation between stellar parameters is similar to what was discussed in Sect~\ref{sect:base}. The hydrogen mass fraction is still the controlling parameter for $Z_0$, $\alpha$ and, subsequently $\stage$. Also of note is that the proportion of shell-burning radiative-core models remains close to what was obtained from runs \#1 and \#2. We have indeed $P(\reps\neq0|\Xv,\rcc=0) = 0.25$ and 0.22 for \run{3} and {4}. Since $P(\rcc=0|\Xv) \approx 1$, these also give the total proportion of early sub-giant models for {\acena}.

\subsection{Microscopic diffusion}

A factor that could reasonably be thought to impact our statistical statements on a convective core presence in {\acena} is microscopic diffusion. The effect we conjecture on here is fairly simple. As diffusion, by means of gravitational settling, diminishes the overall metallicity of the outer convective envelope of {\acena} as time advances, then one will require models with higher initial metallicities to reproduce the surface abundances. This effect is enhanced in the deep interior by the accumulation of heavy elements due to settling. If one neglects overshoot, this in turn might favour, statistically, the onset of convection, as more models may have already entered the CNO regime in early stages, hence with a significant amount of hydrogen left at the centre.

Diffusion is treated in ASTEC using mainly the approximation from \citet{Michaud93}. It provides analytic formula for the various diffusion coefficient entering in Burgers equations \citep{Burgers69}. In particular, under the assumption of a medium composed of fully ionized hydrogen and helium, and after neglecting the trace elements, they obtain an expression for the Coulomb logarithm which in turn enters in the analytic expressions for the collision integrals. Unfortunately, microscopic diffusion for metals is not fully implemented in ASTEC \citep{JCD08a}. It has been included for low-mass stars with a radiative core and a convective envelope, such as the Sun. However, it is not consistently implemented for convective cores, hence limiting considerably the scope of this test. This means that, in order to model {\acena}, we had to consider only the diffusion of helium. It has the following immediate consequence. In a model with gravitational settling of helium relative to a hydrogen-rich background, while settling of heavy elements is neglected, $X$ will increase with age in the convective envelope while $Z$ remains constant. The $Z/X$ ratio will thus decrease. Therefore, in order to reproduce an observed $Z/X$ ratio one could either use lower values of $X_0$ or higher values of $Z_0$. This is to some extent what is seen in Table~\ref{tab:parameters}. The results for \run{5} show that when microscopic diffusion is included the estimated value for $X_0$ is marginally lower and the one for $Z_0$ higher. Following the interpretation given in Sect.~\ref{sect:base} we understand why this does imply a significantly higher value for the mixing-length parameter. Finally the estimated age is very close to what was obtained in runs~\#1 and \#2 for convective-core models. Therefore, it shall come as no surprise that 89\% of the models for {\acena} now display such a feature. 

This large increase in $P(\rcc\neq0|\Xv)$ is {\textquotedblleft}mechanically{\textquotedblright} induced by microscopic diffusion. It does not seem trivial to conjecture on the effect of implementing a full treatment of diffusion that includes metals. However, if metals were to decrease in the convective envelope, then one may have to increase $Z_0$ or decrease $X_0$ even further to reproduce $Z/X$. Therefore, it may not change or even increase the final proportion of convective-core models. However, one may also note that this proportion may be greatly lowered if one were to include LUNA reaction rates. 

\section{Discussion}\label{sect:blabla}

\subsection{Frequency statistical distribution}

We note that the uncertainties on the frequencies given in \citet{Bazot07} remain difficult to interpret. Indeed, they
 have been estimated rather crudely. The authors simply used the frequency resolution of the time series. The rationale behind this was to say that the modes were unresolved (due to the short duration, $T$, of the observing run), and hence could lie anywhere in the interval $[\nu - 1/T, \nu + 1/T]$. In spite of this, Paper~I considered these estimates as representing the standard deviation of a normally distributed random variable (mostly due to algorithmic constraints), hence leading to an inconsistency in the statistical analysis. Of course it is not clear how problematic this really is and how it impacts the finale estimates of the stellar parameters. For this reason, we will also test an alternative formulation of the likelihood
\begin{equation}\label{eq:likeli2}
\pi(\Xv|\thetav) \propto \exp\left[ -\frac{1}{2} \sum_{i=1}^{N_{\mathcal{C}}} \frac{(X_i^{\mathrm{C}} - \mathcal{S}_i^{\mathrm{C}}(\thetav))^2}{\sigma_i^2} \right]\prod_{j=1}^{N_{\mathcal{S}}}T_j(\mathcal{S}_j^{\mathrm{S}}(\thetav);X_j^{\mathrm{S}},\sigma_j).
\end{equation}
Here, $T(x;\mu,\sigma)$ is a symmetrical triangular distribution centred at $\mu$ and with half-width $\sigma$, which is written
\begin{equation}\label{eq:triangular}
\displaystyle
T(x;\mu,\sigma) = \begin{cases}
0& \text{ if $x \leq \mu - \sigma$,}\\
x/\sigma^2 + (\sigma-\mu)/\sigma^2& \text{ if $\mu - \sigma \leq x \leq \mu$,}\\
-x/\sigma^2 + (\sigma+\mu)/\sigma^2& \text{ if $\mu \leq x \leq \mu + \sigma$,}\\
0& \text{ if $\mu + \sigma \leq x$}.
\end{cases}
\end{equation}
 This is the proper description of probability density of the small separations if the frequencies are distributed uniformly \citep[for a general expression see][ the case for two uniform distributions can be straightforwardly obtained from their formula (2.3)]{Bradley02}. The superscripts C and S design respectively the classical and seismic observables.

Table~\ref{tab:parameters} shows the basic results, \run{6}, for a model that equivalent to \run{1} but using a likelihood of the form (\ref{eq:likeli2}). We can see some discrepancies between the results obtained using statistical models (\ref{eq:likeli1}) and (\ref{eq:likeli2}). More specifically, the estimated age is marginally higher in the latter case. This could be due to an intrinsic difference in the formulation of the constraint or to numerical issues. It is indeed difficult to sample from a triangular distribution using Gaussian proposal distribution. Our MCMC algorithm seems to perform slightly less efficiently in this case. These numerical problems can introduce a bias in the final estimates of some parameters. We can also note that the credible interval on the age is slightly smaller, potentially reflecting the tighter constraint provided by (\ref{eq:likeli2}).

The main results however concern the convective onset probability. Here, it has been estimated to $\sim31\%$, which is fairly close to the value obtained using model (\ref{eq:likeli1}). Therefore, we can safely conclude that, even though it would be preferable to have robustly estimated probability densities for the oscillation frequencies of {\acena}, the crude approximation made in Paper~I and here only moderately affect the overall conclusion of our analysis. 

Now that we have established that our conclusions still stand with an alternative statistical model, we can also note that one may expect frequency-separation ratios to be better diagnostics of the stellar interior than small separations. For instance \cite{Miglio05} have argued along this line in order to discard the possibility of overshooting-induced convective cores in {\acena}. However, the tests we made have shown that for the HARPS data at hand, all frequency combinations lead to approximately identical results, due to the large uncertainties.

Finally, we stress the existence of another data set, combining time series from CORALIE and UVES/UCLES \citep{deMeulenaer10}. We ran some MCMC simulations using this data set. However, we found that it was much more difficult to find models fitting it properly. We thus decided that this problem should be treated separately and decided to focus solely on our HARPS data set in this study.

\subsection{Model comparison and nuclear reaction rates}\label{sect:modecomp}

A last point should be discussed, even though it lies at the fringe of the scope of this paper. Given the results of the previous section, a natural question would be whether or not the current data can help us to distinguish between the NACRE and LUNA reaction rates. Since these two formulations lead to very different results to the modelling problem, it is interesting to try to understand if one of them is more accurate. 

A well-known practice in Bayesian Statistics consists in using the tools of marginalization in order to estimate directly model posterior probability instead of model parameter posterior probabilities. 
Let a model\footnote{For instance here: a specific set-up of ASTEC.} be represented by $M$, its posterior probability, the data being given, is (using probability for discrete models rather than continuous densities)
\begin{equation}\label{eq:bayes2}
\displaystyle
P(M|\Xv) = \frac{P(M)P(\Xv|M)}{P(\Xv)}.
\end{equation}
Hence, the ratio of the probability of two competing models, $M_1$ and $M_2$, can be written
\begin{equation}\label{eq:pdfmodratio}
\displaystyle
\frac{P(M_1|\Xv)}{P(M_2|\Xv)} = \frac{P(M_1)}{P(M_2)}\frac{\pi(\Xv|M_1)}{\pi(\Xv|M_2)}.
\end{equation}
This depends on the evidence ratio. By using Eq.~(\ref{eq:bayes}) we obtain immediately
\begin{equation}\label{eq:mlikeli}
\displaystyle
\pi(\Xv|M_i) = \int_{\Theta_i} \pi(\thetav|M_i)\pi(\Xv|\thetav)d\thetav,
\end{equation}
with $\Theta_i$ the space of parameters corresponding to model $i$. This integral is also often called the marginal likelihood, since it is the latter, considered as a function of $\thetav$, that is integrated over the space of parameter, \emph{$\thetav$ being distributed according to the prior density}. This later point is crucial since it explains why performing such integrals can be extremely expensive \citep[see for instance][for a discussion in the astrophysical context]{Ford07}.

Turning back again to Eq.~(\ref{eq:pdfmodratio}), we see that the ratio of the prior probabilities of the models are involved. The second ratio of the right-hand side, usually called the Bayes factor, is often used as the starting point for Occam-factor-type arguments aiming at distinguish models $M_1$ and $M_2$. It has however been argued by \citet{Wolpert92}, in the context of supervised machine learning, that such a procedure is not straightforward and even flawed in many aspects. Therefore great caution is necessary when trying to make inference about models using the evidence. 

An interesting example in our case would consist in determining which formulation of the $^{14}$N$(p,\gamma)^{15}$O nuclear reaction rate is in better agreement with our data. In Eq.~(\ref{eq:pdfmodratio}), let us set consider $M_1 = M_{\mathrm{LUNA}}$ (i.e. ASTEC with the LUNA reaction rate included for $^{14}$N$(p,\gamma)^{15}$O) and   $M_1 = M_{\mathrm{NACRE}}$ (only NACRE reaction rates in ASTEC). In the limited scope of this paper, we can only make the simple assumption that the two models are \emph{a priori} equally probable and set $p(M_{\mathrm{LUNA}}) = p(M_{\mathrm{NACRE}})$ \citep[the {\textquotedblleft}principle of indifference{\textquotedblright}, see][]{Loredo90}. When this is done, it is possible to compute the evidence of each model. The method we used is described in Appendix~\ref{app:cj01}.

Using the above assumption on $p(M_{\mathrm{LUNA}})$ and $p(M_{\mathrm{NACRE}})$ and Eq.~(\ref{eq:chib}), we find an estimate of $\log[p(\Xv|M_{\mathrm{NACRE}})/P(\Xv|M_{\mathrm{LUNA}})] = \log[p(M_{\mathrm{NACRE}}|\Xv)/P(M_{\mathrm{LUNA}}|\Xv)] = 0.33$. If we use a qualitative classification \citep[see e.g.][]{Robert07}, it is not possible to distinguish one model from the other and thus one would conclude that the data are insufficient to discriminate between the different nuclear reaction rates for $^{14}$N(p,$\gamma$)$^{15}$O found in the literature. However, such an odds ratio could conceivably be balanced by a different prior ratio favouring the LUNA reaction rate, the limiting cases being  $p(M_{\mathrm{LUNA}}) = 1$ and $p(M_{\mathrm{NACRE}}) = 0$ if the LUNA reaction rates are believed to be better than the NACRE ones, and conversely.

The only way to obtain such information is by using results coming from nuclear reaction experiments, from a statistical comparison between the NACRE and LUNA setups. In both cases, equal or unequal priors, the data might still show a better agreement with one of the models. Yet, only with statistically robust priors, reflecting our information (the two experiments are not equivalent and neither are the subsequent models), can the final posterior probability ratio be properly estimated. For the time being, since such comparison experiments between reaction-rate measurements do not exist, to the extent of our knowledge, quantifying a prior ratio to include in Eq.~(\ref{eq:pdfmodratio}) might represent a logical continuation to our analysis in stellar physics.

\subsection{Impact of non-equilibrium abundances}\label{sect:neq}

In their study of the star HD~203608, a low-mass main-sequence star with a convective core, \citet{Deheuvels10} discussed the effect of non-equilibrium $^3$He and $^7$Li abundances on nuclear energy generation. In particular, they focused on how these can sustain a convective core even with dominant ppI or ppII cycles. 

Their main argument goes as follow. The ppI and ppII cycles do not generate high enough energy generation rates to sustain a convective core. However, when $^3$He or $^7$Li abundances are above their equilibrium values, then the temperature sensitivities of, respectively, the $^3$He($^3$He,$\gamma$)$^4$He and $^7$Li(p,$^4$He)$^4$He reactions increase and may lead to a convective onset or sustain an already existing convective core. This core would last at least until the abundances reach their equilibrium values. In case overshooting is included, the additional inflow of $^3$He or $^7$Li can increase the lifetime of a convective core for a significant amount of time. 

As already mentioned in Sect.~\ref{sect:ASTEC}, out-of-equilibrium abundances for $^3$He or $^7$Li were not used in ASTEC in this study\footnote{As of today, an out-of-equilibrium treatment of $^3$He is available in ASTEC. However, it leads to numerical instabilities in convective cores and was thus not suited to an estimation of the stellar parameters of {\acena} based on an MCMC algorithm.}. This certainly raises questions about the robustness of our results. We cannot proceed here to a full-scale numerical verification using MCMC simulations, which would require the implementation of the said non-equilibrium effects and should be addressed in a future study. However, it is possible to give qualitative assessments on the validity of our results.

\begin{figure}
\center
\includegraphics[width=\columnwidth]{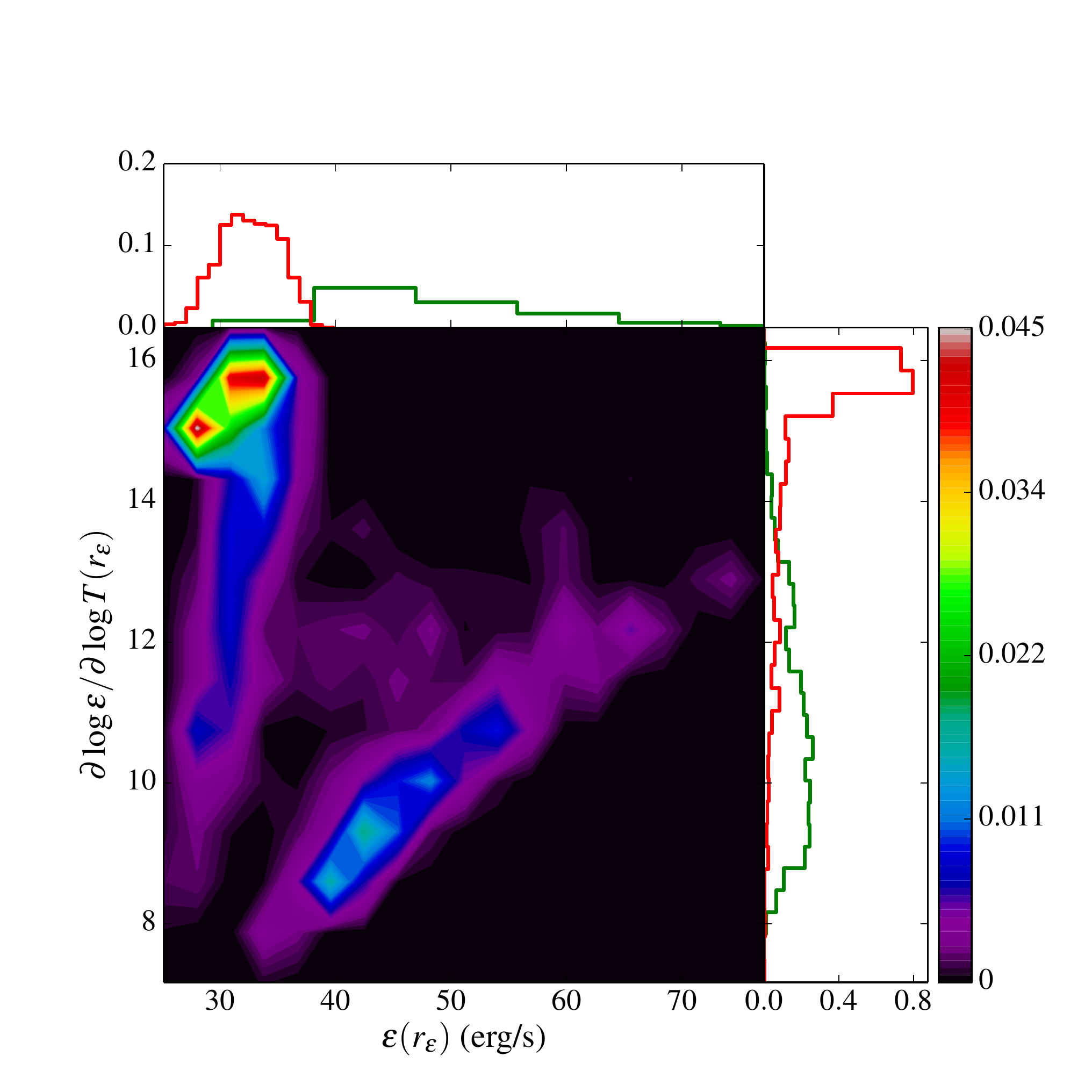}
\caption{Joint PDF for $\varepsilon(\reps)$ and $\partial\log\varepsilon/\partial\log T (\reps)$, with $\reps$ the radius of maximum energy generation rate as defined in Sect.~\ref{sect:base}.}
\label{fig:sensitivity}
\end{figure}

If no overshoot is included, our simplifications should hardly make any difference. If it is taken into account, then we can distinguish two cases, depending on whether or not it has a convective core. In the former case, convection is most likely sustained by the ppII cycle. According to our results, an out-of-equilibrium $^7$Li(p,$^4$He)$^4$He reaction, with an increased temperature sensitivity, is not a necessary condition for central convection to exist. An influx of H, increasing the energy generation rate without changing much the temperature sensitivity, is clearly enough. We illustrate this point in Fig.~\ref{fig:sensitivity} in which we plot the joint PDF for ($\varepsilon(\reps),\partial\log\varepsilon/\partial\log T (\reps)$). This however does not imply that out-of-equilibrium $^7$Li abundances are not a sufficient condition. It remains to be seen in which proportions these two effects interact when treated simultaneously. This is an interesting element of discussion for future research. Moreover, we also noted that the CNO cycle, which contributes roughly a third of the energy generation rates in convective models of {\acena}, was increasing the temperature sensitivity of $\ec$. If out-of-equilibrium $^7$Li abundances can produce the same effect, then the relative importance of CNO could be further diminished.

If the model has no convective core, we saw that CNO is then very likely to be the main source of energy production in the central regions of {\acena}. Out-of-equilibrium $^3$He or $^7$Li may thus not be of critical importance at this stage. A possibility for these to matter is if we {\textquotedblleft}missed{\textquotedblright} convective cores (already disappeared at the final age of {\acena}) that would have been sustained during the main sequence evolution due to a overshooting maintaining ppI or ppII cycles out of equilibrium. However, it is unlikely that such unseen convective cores would induce structural effects large enough during the main-sequence evolution to modify significantly, and as a whole, the subset of models without convective core. This assumption is somewhat supported by Sect. 3.3 of \citet{Deheuvels10}. 

\section{Conclusion}

We have analysed {\acena} using interferometric, spectroscopic, astrometric and seismic constraints. We modelled the star using the ASTEC code. In order to estimate the stellar parameters, we chose a Bayesian framework, allowing us to make probabilistic inferences about these quantities. We used some of the most recent equations of state and opacities. We also considered the effect of changes in the nuclear reaction rates and of including overshoot or microscopic diffusion.

We obtained realistic estimates of the stellar parameters, in the sense that they reflect our current knowledge of the physical state of {\acena}, given the data. We also obtained a fairly contrasted and somewhat complex picture of the star. First, our conclusions do not not seem, when compared to the previous results of Paper~I, to depend strongly on the precise formulation of the equation of state or the opacity. On the other hand, nuclear reaction rates and microscopic diffusion appear to be major sources of uncertainty for the modelling. 

A great emphasis was put on the nature of energy transport in the central regions of the star, as well as its generation. For models using the NACRE reaction rates, there is almost a 40\% chance for {\acena} to have a convective core. Its physical properties depend on the details of the physics included. In particular, overshoot leads to convective cores with fractional radii of the order of 0.08. This roughly is twice the size of the convective cores observed for models without overshoot. Bringing more hydrogen to the central region, it can also favour energy generation processes such as the ppII chain. We have found that in some cases the temperature sensitivity of the ppII chain contributes substantially to the onset of convective instability in the core. However, the CNO cycle remains a major component of a potential convective onset. Switching to the LUNA reaction rates, which decrease strongly the efficiency of the CNO cycle, we saw convective-core models almost completely disappear from our solutions. Finally, microscopic diffusion might increase the probability of obtaining a convective core in {\acena}.

It thus appears that the complexity of the picture we are drawing has its roots in the uncertainties on the physical processes at play. This is particularly sensitive because {\acena} lies in a transition region regarding energy transport, and each minor change in our model may have strong repercussions on the central structure of the star. There is thus ample work remaining in order to select the proper physics for this star. A clear line of investigation for the future will consist in clarifying the role played by physical process such as non-equilibrium nuclear reactions. This could represent a serious alternative to our interpretation of convective cores sustained by the ppII cycle. It even has the potential to change the overall complexion of our results. We shall also consider the effects of additional mixing, either driven by rotation, gravity waves or other processes that could compete with overshoot, in order to better understand the values of $\alphaov$ estimated here.

This has to be done partially through the use of new observational constraints, using for instance the marginal likelihood approach described above. With that in mind, one may wish for higher-quality observational data, in particular seismic. Long, high-duty cycle, high signal-to-noise ratio spectroscopic time series are desirable in order to improve the existing seismic data. We saw that the convective and radiative-core model populations (when existing in the solutions) have somewhat different properties. More precise oscillation frequencies, if they can indeed rule out one of the two options, might help to improve the precision on the stellar parameters, which is sometimes low, in particular for the age.

However, just improving the observations will not provide all the answers we seek. For instance, when using the marginal likelihood approach, one also needs to make assessments on the prior probabilities. This can only be achieved by comparing, through experimentation or maybe numerical simulations, the different physical processes at play (nuclear reaction rates, microscopic diffusion,\dots) and their suggested formulations.

We also showed that, almost independently of the selected physics (with the notable exception of microscopic diffusion), a significant fraction of models seem to have already started their turn on the subgiant branch. In general we find 25\% to 30\%  of the models for which the peak of energy generation is not at the centre. That being said, these nearly isothermal cores are never very large (only a few percents of the stellar radius), meaning that here too {\acena} might lie in a transition region. In short, by looking as carefully as we could at the physical state of {\acena}, we discovered that there remain physical uncertainties, which translate into challenges either in terms of modelling or future observational campaigns.

\section*{Acknowledgements}

We would like to thank M.~J.~P.~F.~G.~Monteiro for enduring kindly the tentative interpretations of MB. Micha\"el Bazot would like to thank S.~Hannestad for providing him access to the Grendel cluster at DCSC/AU of which important use has been made during this work. This material is based upon work supported by the NYU Abu Dhabi Institute under grant G1502. Funding for the Stellar Astrophysics Centre is provided by The Danish National Research Foundation (Grant DNRF106). The research is supported by the ASTERISK project (ASTERoseismic Investigations with SONG and Kepler) funded by the European Research Council (Grant agreement no.: 267864). Finally, the authors would like to thank the anonymous referee for his/her very constructive remarks.

\appendix

\section{Chib \& Jezliakov algorithm}\label{app:cj01}

For the sake of completeness, we give here the outline of the algorithm suggested by \citet{Chib01} to estimate the marginal likelihood (\ref{eq:mlikeli}). A relative straightforward way to do so is to rewrite Bayes' formula 
\begin{multline}\label{eq:mlikeli2}
\log \pi(\Xv|M_i) = \\
\log \pi(\Xv|M_i,\thetav^{\ast}) + \log\pi(\thetav^{\ast}|M_i) - \log\widehat{\pi}(\thetav^{\ast}|M_i,\Xv).
\end{multline}
Here the right-hand side is independent of $\thetav$ and $\thetav^{\ast}$ is a suitably chosen value for the parameters, usually chosen in a region of high probability density to facilitate the sampling. The term $\widehat{\pi}(\thetav^{\ast}|M_i,\Xv)$ is an estimate of the PDF\footnote{Note here that the MCMC algorithm does not provide us with this value, it just generates a sample distributed according to the PDF. For parameter estimation purposes, we do not need the normalization constant of $\pi(\thetav|\Xv,M_i)$ in order to compute moments or find maxima.} given by the formula \citep{Chib01}
\begin{equation}\label{eq:chib}
\displaystyle 
\widehat{\pi}(\thetav^{\ast}|\Xv,M_i) = \frac{\Exp^{\pi(.|\Xv,M_i)}[\rho(\thetav,\thetav^{\ast}|\Xv)q(\thetav,\thetav^{\ast}|\Xv)]}{\Exp^{q(.,\thetav^{\ast}|\Xv)}[\rho(\thetav,\thetav^{\ast}|\Xv)]}.
\end{equation}
Here $\Exp^{\pi(.|\Xv,M_i)}$ is the expectation value with respect to $\pi(.|\Xv,M_i)$ and can thus be computed thanks to the output of the MCMC simulation, using $(1/M)\sum_{j=1}^M \rho(\thetav^{(j)},\thetav^{\ast}|\Xv)q(\thetav^{(j)},\thetav^{\ast}|\Xv)$, with $\{ \thetav^{(1)},\dots,\thetav^{(M)}\} \sim \pi(.|\Xv,M_i)$. The term in the denominator is the expectation with respect to $q(.,\thetav^{\ast})$ and can be estimated using $(1/N)\sum_{i=1}^N \rho(\thetav^{(i)},\thetav^{\ast}|\Xv)$, with $\{ \thetav^{(1)},\dots,\thetav^{(N)}\} \sim q(.,\thetav^{\ast})$. Note that this requires the computation of $N$ additional models. Furthermore, the formula is obviously only valid for homogeneous Markov chains, meaning that $\rho(\thetav^{(t)},\thetav^{\ast})$ cannot depend on the rank $t$ of the MCMC iteration. This is of course not verified in the AMCMC framework during adaption phases. Consequently one shall work using a subsample of the MCMC sample for which $q(.,\thetav^{\ast})$ does not change.

\bibliography{MCMCref}

\end{document}